\documentclass[10pt,twocolumn,letterpaper]{article}

\usepackage{cvpr} 

\usepackage{graphicx}
\usepackage{amsmath}
\usepackage{amssymb}
\usepackage{booktabs}
\usepackage{float}
\usepackage{stfloats}
\usepackage{multirow}
\usepackage{diagbox}
\usepackage{makecell}
\usepackage[pagebackref,breaklinks,colorlinks]{hyperref}
\usepackage[accsupp]{axessibility}
\usepackage[toc,page]{appendix}
\usepackage[capitalize]{cleveref}
\crefname{section}{Sec.}{Secs.}
\Crefname{section}{Section}{Sections}
\Crefname{table}{Table}{Tables}
\crefname{table}{Tab.}{Tabs.}

\def\cvprPaperID{6090} 
\def\confName{CVPR}
\def\confYear{2023}

\begin{document}

\title{Neural Video Compression with Diverse Contexts}

\author{Jiahao Li, Bin Li, Yan Lu\\
Microsoft Research Asia\\
{\tt\small \{li.jiahao, libin, yanlu\}@microsoft.com}
}
\maketitle

\begin{abstract}
For any video codecs, the coding efficiency highly relies on whether the current signal to be encoded can find the relevant contexts from the previous reconstructed signals. Traditional codec has verified more contexts bring substantial coding gain, but in a time-consuming manner. However, for the emerging neural video codec (NVC), its contexts are still limited, leading to low compression ratio. To boost NVC, this paper proposes increasing the context diversity in both temporal and spatial dimensions. First, we guide the model to learn hierarchical quality patterns across frames, which enriches long-term and yet high-quality temporal contexts. Furthermore, to tap the potential of optical flow-based coding framework, we introduce a group-based offset diversity where the cross-group interaction is proposed for better context mining. In addition, this paper also adopts a quadtree-based partition to increase spatial context diversity when encoding the latent representation in parallel. Experiments show that our codec obtains 23.5\% bitrate saving over previous SOTA NVC. Better yet, our codec has surpassed the under-developing next generation traditional codec/ECM in both RGB and YUV420 colorspaces, in terms of PSNR. The codes are at \url{https://github.com/microsoft/DCVC}.

\end{abstract}

\section{Introduction}
\label{sec:intro}
The philosophy of video codec is that, for the current signal to be encoded, the codec will find the relevant contexts (e.g., various predictions as the contexts)  from previous reconstructed signals to reduce the spatial-temporal redundancy. The more relevant contexts are, the higher bitrate saving is achieved.

If looking back the development of traditional codecs (from H.261 \cite{girod1995comparison}  in 1988 to H.266 \cite{bross2021overview} in 2020), we find that the coding gain mainly comes from the continuously expanded coding modes, where each  mode uses a specific manner to extract and utilize context.  For example, the numbers of intra prediction directions \cite{pfaff2021intra} in H.264, H.265, H.266 are 9, 35, and 65, respectively. 
So many modes can extract diverse contexts to reduce redundancy, but also bring huge complexity as rate distortion optimization (RDO) is used to search the best mode. For encoding a 1080p frame, the under-developing ECM (the prototype of next generation traditional codec) needs up to half an hour \cite{ECM_performance}. Although some DL-based methods \cite{su2019machine,su2019machine2, huang2019cnn} proposed accelerating traditional codecs, the complexity is still very high. 

By contrast, neural video codec (NVC) changes the extraction and utilization of context from hand-crafted design to automatic-learned manner. Mainstream frameworks of NVC can be classified into residual coding-based \cite{lu2019dvc,Rippel_2021_ICCV,lu2020end,lin2020m,agustsson2020scale,Djelouah_2019_ICCV,yang2021learning,wu2018video,liu2020neural} and condition coding-based \cite{liu2020conditional, ladune2021conditional,canfvc,vct, li2021deep,sheng2021temporal, li2022hybrid}. The residual coding explicitly uses the predicted frame as the context, and the context utilization is restricted to use subtraction for redundancy removal. By comparison, conditional coding implicitly learns feature domain contexts. The high dimension contexts can carry richer information to facilitate encoding, decoding, as well as entropy modelling.

 \begin{figure}[t]
		\begin{center}
			\includegraphics[width=1\linewidth]{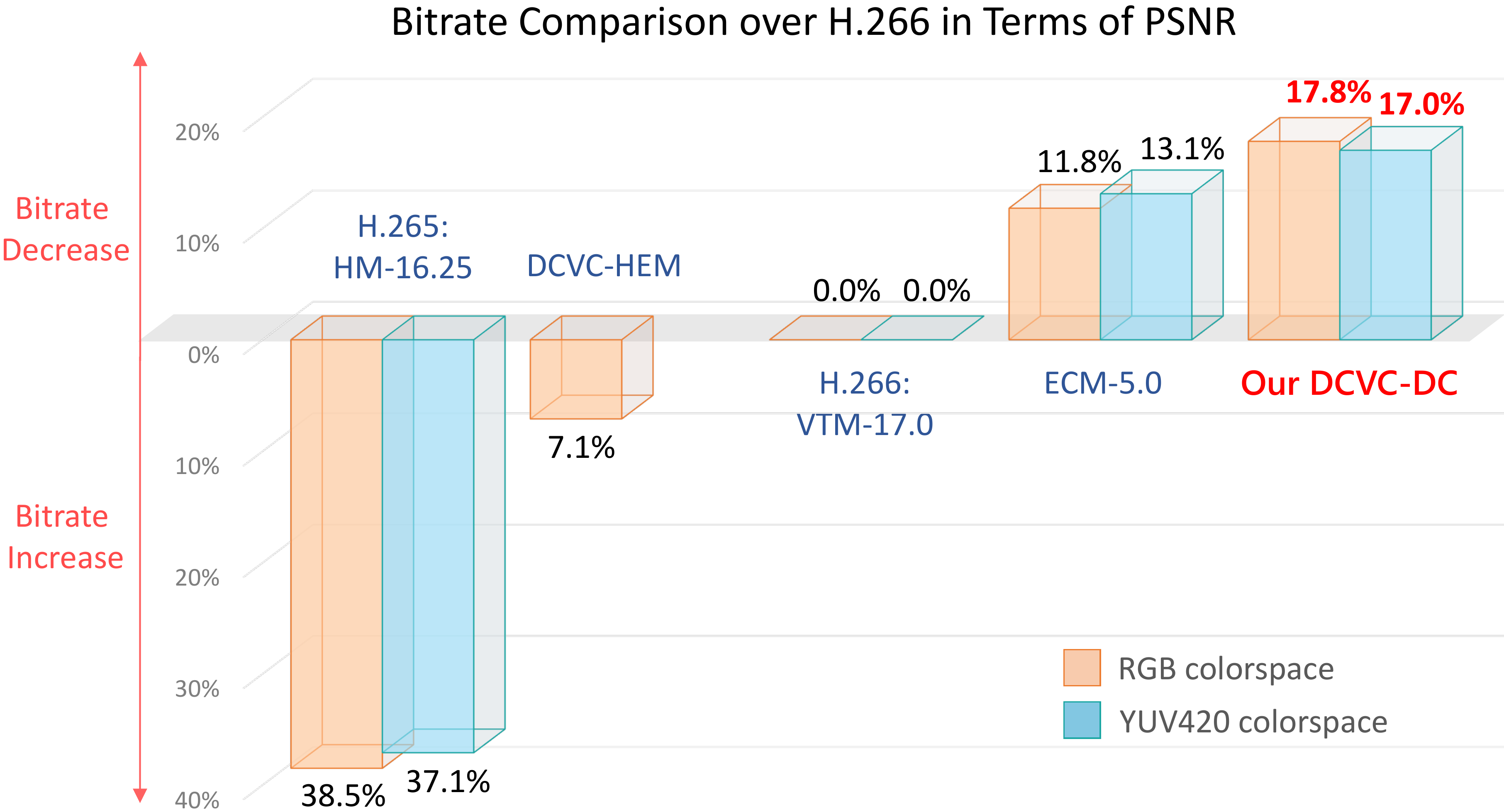}
		\end{center}
		\vspace{-0.2cm}
		\caption{Average results on UVG, MCL-JCV, and HEVC datasets. All traditional codecs use their  best compression ratio configuration. DCVC-HEM\cite{li2022hybrid} is the previous SOTA NVC and only has released the model for RGB colorspace.		}
		\vspace{-3mm}
		\label{PerformanceSummary}
	\end{figure}
However, for most NVCs, the manners of context extraction and utilization are still limited, e.g., only using  optical flow to explore temporal correlation. This makes NVC  easily suffer from  the  uncertainty \cite{ma2021uncertainty, der2009aleatory, Gal2016Uncertainty} in  parameters or fall into local optimum \cite{huo2022towards}. One solution is adding traditional codec-like coding modes into NVC \cite{huo2022towards}. 
But it brings large computational complexity as RDO is used.   
So the question comes up: how to better learn and use the contexts while yielding low computational cost?
 
To this end, based on DCVC (deep contextual video compression) \cite{li2021deep} framework and its following work DCVC-HEM \cite{li2022hybrid}, we propose a new model DCVC-DC which efficiently utilizes the Diverse Contexts to further boost compression ratio. At first, 
we guide DCVC-DC to learn hierarchical quality pattern across frames. 
With this guidance during the training, the long-term and yet high-quality contexts which are
vital for the reconstruction of the following frames are implicitly learned during the feature propagation.
 This helps further exploit the long-range temporal correlation in video and effectively alleviate the quality degradation problem existed in most NVCs.
 In addition, we adopt the  offset diversity \cite{ODPaper} to strengthen the optical flow-based codec, where multiple offsets can reduce the warping errors for complex or large motions. In particular, inspired by the weighted prediction in traditional codec, the offsets are divided into groups and the cross-group fusion is proposed to improve the temporal context mining. 

Besides from temporal dimension, this paper also proposes increasing the spatial context diversity  when encoding the latent representation. Based on recent checkerboard model \cite{he2021checkerboard} and dual spatial model \cite{li2022hybrid, wang2023EVC}, we design a quadtree-based partition to improve the distribution estimation.
When compared with \cite{he2021checkerboard, li2022hybrid}, the types of correlation modelling are more diverse 
hence the model has a larger chance to find more relevant context. 

It is noted that all our designs  are parallel-efficient. To further reduce the computational cost, we also adopt depthwise separable convolution \cite{chollet2017xception}, and assign unequal channel numbers for features with different resolutions. Experiments  show that our DCVC-DC  achieves much higher efficiency over previous SOTA NVC and pushes the compression ratio to a new height. When compared with DCVC-HEM \cite{li2022hybrid}, 23.5 \% bitrate saving is achieved while  
 MACs (multiply–accumulate operations) are reduced by 19.4\%.
 Better yet, besides H.266-VTM 17.0, our codec also already outperforms ECM-5.0 (its best compression ratio configuration for low delay coding is used) in both RGB and YUV420 colorspaces, as shown in Fig. \ref{PerformanceSummary}. To the best of our knowledge, this is the first NVC which can achieve such accomplishment.
 In summary, our contributions are:

\begin{itemize}
    \item We propose efficiently increasing context diversity to boost NVC. Diverse contexts are complementary to each other and have larger chance to provide good reference for reducing redundancy. 
    \item From temporal dimension, we guide model to extract high-quality contexts to alleviate the quality degradation problem. In addition, the group-based offset diversity is designed for better temporal context mining.
    \item From spatial dimension, we adopt a quadtree-based partition for latent representation. This provides diverse spatial contexts  for better entropy coding.
    \item Our DCVC-DC obtains 23.5\% bitrate saving over the previous SOTA NVC. In particular, our DCVC-DC has surpassed the best traditional codec ECM in both RGB and YUV420 colorspaces, which is an important milestone in the development of NVC.
\end{itemize}

\section{Related Work}
\label{sec_related_work}
\subsection{Neural Image Compression}
Most neural image codecs are based on hyperprior \cite{balle2018variational} where some  bits are first used to  provide basic  contexts for entropy coding. Then, the auto-regressive prior \cite{minnen2018joint} proposes using neighbour contexts to capture spatial correlation. 
Recently works \cite{qian2020learning,kim2022joint, guo2021causal, qian2022entroformer} propose extracting global or long-range contexts to further boost performance. These show more diverse contexts bring substantial coding gain for neural image codec.

\subsection{Neural Video Compression}
Recent years also have witnessed the prosperity of NVC. The pioneering  DVC \cite{lu2019dvc} follows traditional codec. It uses optical flow network to generate  prediction frame, then its residual with the current frame is coded. Many subsequent works also adopt this residual coding-based framework and refine the modules therein. For example, \cite{lin2020m,Rippel_2021_ICCV, pourreza2022boosting} proposed motion prediction to further reduce redundancy. Optical flow estimation in scale-space \cite{agustsson2020scale} was designed to handle complex motion. Yang \textit{et al.} \cite{yang2021learning} utilized recurrent auto-encoder to improve coding efficiency. 

 Residual coding explicitly generates predicted frame in pixel domain  as the context and only uses the subtraction to remove redundancy. By comparison, conditional coding has stronger extensibility. The definition, learning, and usage manner of condition can be  flexibly designed. In \cite{liu2020conditional,vct}, temporally conditional entropy models were designed. 
 \cite{ladune2021conditional} uses conditional coding to encode the foreground contents. Li \textit{et al.} proposed DCVC \cite{li2021deep}  to learn feature domain contexts to increase context capacity. Then DCVC-TCM \cite{sheng2021temporal} adopts feature propagation  to  boost performance. Recently, DCVC-HEM \cite{li2022hybrid} designs the hybrid entropy model utilizing both spatial and temporal contexts. 

However, the coding modes in most NVCs are still limited when compared with traditional codec. For example, traditional codec adopts translational/affine motion models, geometric partition, bi-prediction, and so on modes to extract diverse temporal contexts \cite{bross2021overview}. By contrast, existing NVCs usually only relies on single optical flow, which is easily influenced  by epistemic uncertainty \cite{ma2021uncertainty, der2009aleatory, Gal2016Uncertainty} 
in model parameters. The recent work \cite{huo2022towards} also shows such NVCs easily fall into local optimum when coding mode is limited. So \cite{huo2022towards} designed many additional modes like traditional codec, and   RDO is used to search the best mode. Such method inevitably brings large computational cost. By comparison, our model has no additional inference cost when exploiting high-quality temporal contexts. Our offset diversity and quadtree  partition are also time-efficient designs in providing diverse contexts.

\section{Proposed Method}

 \begin{figure*}[t]
		\begin{center}
			\includegraphics[width=1\linewidth]{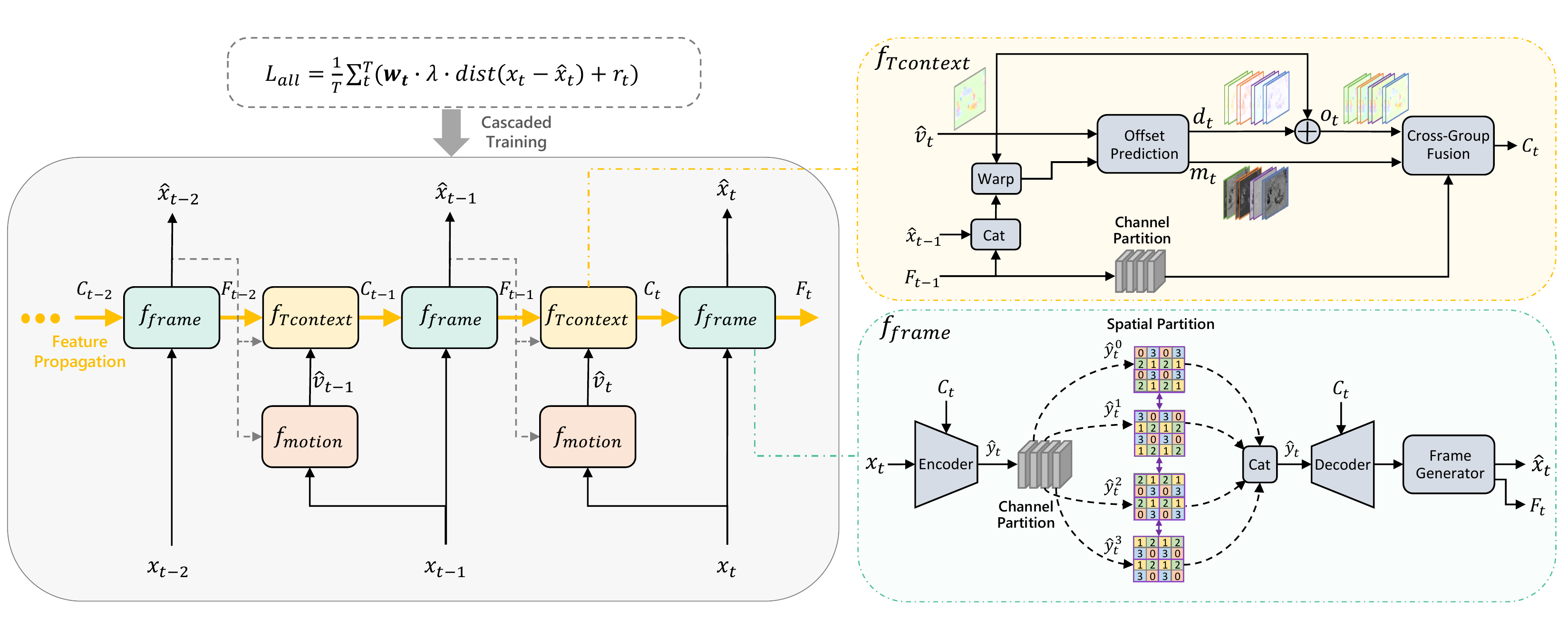}
		\end{center}
		\vspace{-0.5cm}
		\caption{Framework overview of our DCVC-DC. $x_t$ and $\hat{x}_{t}$ are the input and reconstructed frames. $C_t$ is the learned temporal context as the condition for coding  $x_t$. $F_t$ is the propagated but unprocessed feature used for next frame. 	In the loss term, $r_t$ means the bit cost for coding whole frame. $dist(\cdot)$ is distortion function. $\lambda$ and $w_t$ are the global and frame-level weights, respectively. The number in the spatial partition for $\hat{y}_{t}$ represent the coding order index. 	}
		\vspace{-3mm}
		\label{framework}
		
	\end{figure*}
	    
\subsection{Overview}

To achieve higher compression ratio, our codec is built on the more flexible conditional coding rather than the residual coding. The framework of our DCVC-DC is illustrated in Fig. \ref{framework}. It is noted our DCVC-DC is designed for low-delay coding as it can be applied in more scenarios, e.g., real-time communication. As shown in Fig. \ref{framework}, for coding each frame $x_t$  with frame index $t$, our coding pipeline contains three core steps: $f_{motion}$,  $f_{Tcontext}$, and $f_{frame}$. At first, $f_{motion}$ uses  optical flow network to estimate the motion vector (MV) $v_t$, then $v_t$ is encoded and decoded as  $\hat{v}_{t}$. Second, based on $\hat{v}_{t}$ and the propagated feature  $F_{t-1}$ from the previous frame, $f_{Tcontext}$ extracts the motion-aligned temporal context feature $C_{t}$. At last, conditioned on $C_{t}$, $f_{frame}$ encodes $x_t$ into quantized latent representation $\hat{y}_t$. After  entropy coding, the output frame $\hat{x}_{t}$ is reconstructed via the decoder and frame generator. At the same time, $F_{t}$ is also generated and propagated to the next frame. It is noted that, 
our DCVC-DC is based on DCVC-HEM \cite{li2022hybrid}. When compared with  DCVC-HEM, this paper redesigns the modules to exploit Diverse Contexts from both temporal (Section \ref{hierarchical_quality} and \ref{offset_diversity}) and spatial (Section \ref{quadtree}) dimensions. 

\subsection{Hierarchical Quality Structure}

\begin{figure}[t]
		\begin{center}
			\includegraphics[width=\linewidth]{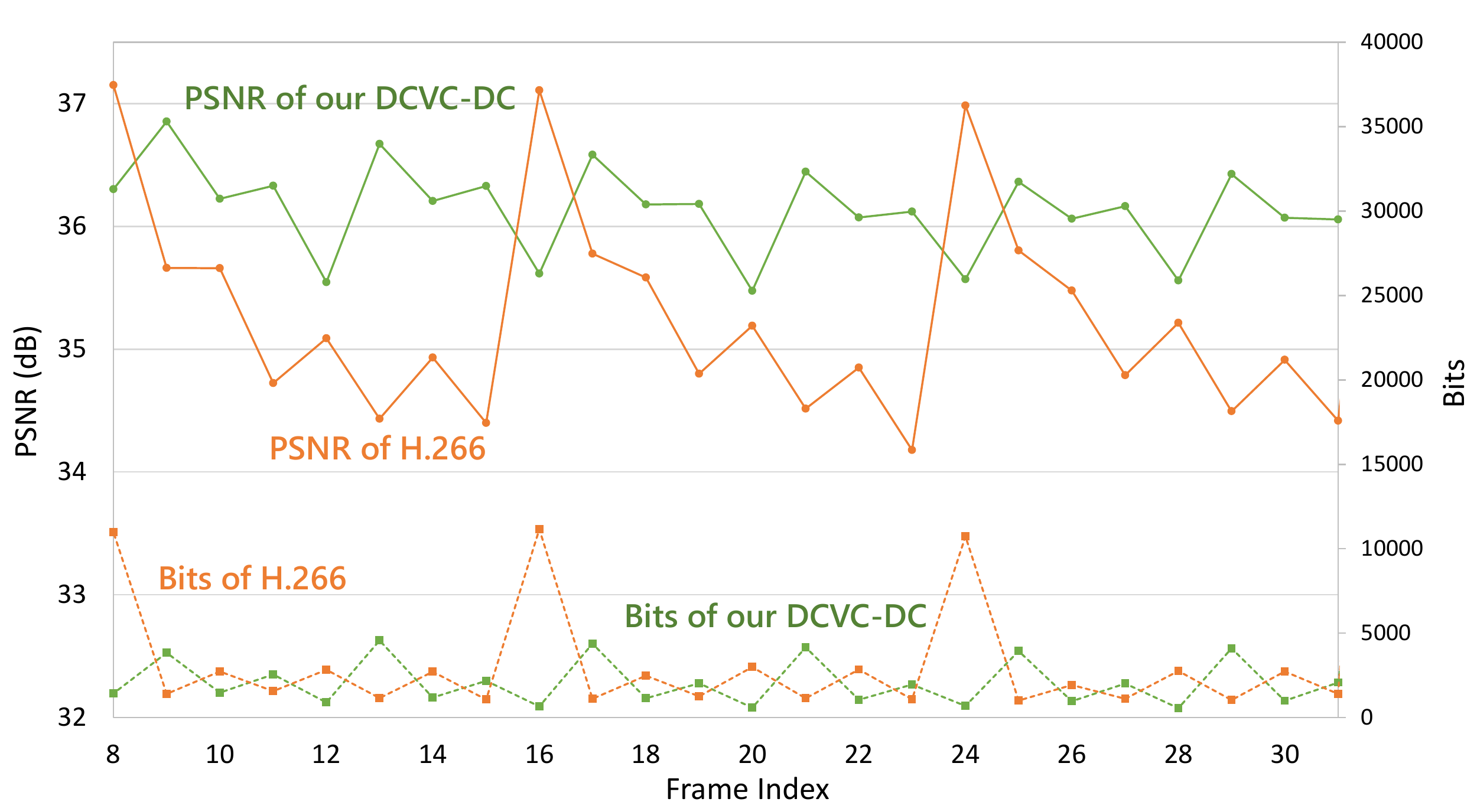}
		\end{center}
		\vspace{-0.6cm}
		\caption{Hierarchical quality structure in H.266-VTM 17.0 and our NVC. This example is from \textit{BasketballPass} video from HEVC D dataset. The average bpp (bits per pixel) and PSNR of H.266 are 0.056 and 35.10. Our DCVC-DC is with 0.045 and 36.13.	}
		\vspace{-6mm}
		\label{HieQP}
	\end{figure}

Traditional codec widely adopts hierarchical quality structure, where frames are assigned into different layers and then use different QPs (quantization parameters). 
This design originates from  scalable video coding
\cite{schwarz2006analysis} 
but also improves the performance for general low delay coding from two aspects. One is  periodically improving the quality can alleviate the error propagation, as shown in Fig. \ref{HieQP}. During the inter prediction, the high-quality reference frames enable the codec to find the more accurate MV during motion estimation. Meanwhile, the motion compensated prediction is also high-quality, leading to smaller prediction error. Another aspect is that, powered by multiple reference frame selection and weighted prediction mechanisms,  the prediction combinations from the nearest  reference frame and long-range high-quality reference frame are more diverse. The work \cite{lim2011reference} investigates many settings on frame quality  and reference frame selection, and concludes that the hierarchical quality structure with referencing both the nearest frame and farther high-quality  frame  achieves the best performance. 

Inspired by the success in traditional codec, we are thinking whether we can equip NVC with the hierarchical quality structure and let NVC also enjoy the benefits. Considering the recent neural codecs \cite{li2022hybrid, cui2021asymmetric} also support variable bitrate in single model, one straightforward solution is following the traditional codec and  directly assigning hierarchical QPs during the inference of NVC. However, not like traditional codec uses well-defined rules to perform motion estimation and motion compensation (MEMC), NVC uses neural networks and MEMC is often in feature domain. For NVC, the advantage of such design is that it is automatic learned and 
has larger potential to achieve better performance. The  disadvantage is that it has wore robustness and  generalization ability for out-of-distribution quality pattern.
Thus, if we directly feed NVC the hierarchical QPs during the testing like \cite{hu2022coarse}, it may not well adapt to the hierarchical quality pattern, and the MEMC may get sub-optimal performance.

\label{hierarchical_quality}
	
To this end, we propose guiding the NVC to learn the hierarchical quality pattern during the training. Specifically, we add a weight $w_t$ for each frame in the rate-distortion loss, as shown in Fig. \ref{framework}. The setting of $w_t$ follows the hierarchical structure. 
Powered by this hierarchical distortion loss, both high-quality output frame   $\hat{x}_{t}$ and feature $F_{t}$ containing many high-frequency details are periodically generated.  They are very helpful for improving the MEMC effectiveness and then alleviate the error propagation problem that many other NVCs suffer from. In addition, via the cascaded training across multiple frames, the feature propagation chain is formed. The high-quality contexts which are vital for the reconstruction of the following frames  are automatically learned and kept in long range. Thus, for the encoding of $x_{t}$, $F_{t-1}$ not only contains the short-term contexts extracted from $x_{t-1}$, but also provides long-term and continually-updated high-quality contexts from many previous frames. Such diverse $F_{t-1}$ helps further exploit the temporal correlation across many frames and then boost the compression ratio. Fig. \ref{HieQP} also shows the quality pattern of our NVC. We can see that our DCVC-DC achieves better average quality while with smaller bit cost than H.266.


\subsection{Group-Based Offset Diversity}

Due to the various motions between frames, directly using the unprocessed $F_{t-1}$ without motion alignment is hard for codec to capture temporal correspondence. Therefore, we  follow existing NVCs   and use optical flow network to extract motion aligned temporal context $C_t$ via  $f_{Tcontext}(F_{t-1}, \hat{v}_{t} )$, where $\hat{v}_{t}$ is the decoded MV.  However, in most existing NVCs, $f_{Tcontext}$ is only the warping operation with the single MV. Such single motion-based alignment 
is not robust to complex motions or occlusions. 
The  works \cite{tian2020tdan, wang2019edvr, chan2022basicvsr++} show deformable alignment gets better results for video restoration as each location has multiple offsets to capture temporal correspondence. So recently it is also applied into NVC \cite{hu2021fvc}. However, the training of deformable alignment is not stable and the overflow of offsets degrades the performance \cite{wang2019edvr,ODPaper}. In addition, the number of offsets is limited to the size of deformable convolutional kernel.
Thus,  this paper adopts the more flexible design called offset diversity \cite{ODPaper}. Meanwhile, the decoded MV is used as the base offset to stabilize the training as \cite{chan2022basicvsr++}. 

As shown in Fig. \ref{framework}, our $f_{Tcontext}$ consists of two core sub-modules: offset prediction and cross-group fusion.  At first, offset prediction uses the decoded MV $\hat{v}_{t}$ to predict the residual offsets $d_t$, where  $\hat{x}_{t-1}$ and $F_{t-1}$ are also warped and  fed as the auxiliary information. The $d_t$ adds the base offset $\hat{v}_{t}$ to obtain the final offsets $o_t$. At the same time, offset prediction also generates the modulation mask $m_t$ which can be regarded as an attention that reflects the confidence of offsets. It is noted that  $F_{t-1}$ is divided into $G$ groups along the channel dimension, and each  group has separate $N$ offsets. Thus, there are a total of $G\times N$ offsets learned. The diverse offsets are complementary to each other, and help codec cope with complex motion and occlusion.

\begin{figure}[t]
		\begin{center}
			\includegraphics[width=1\linewidth]{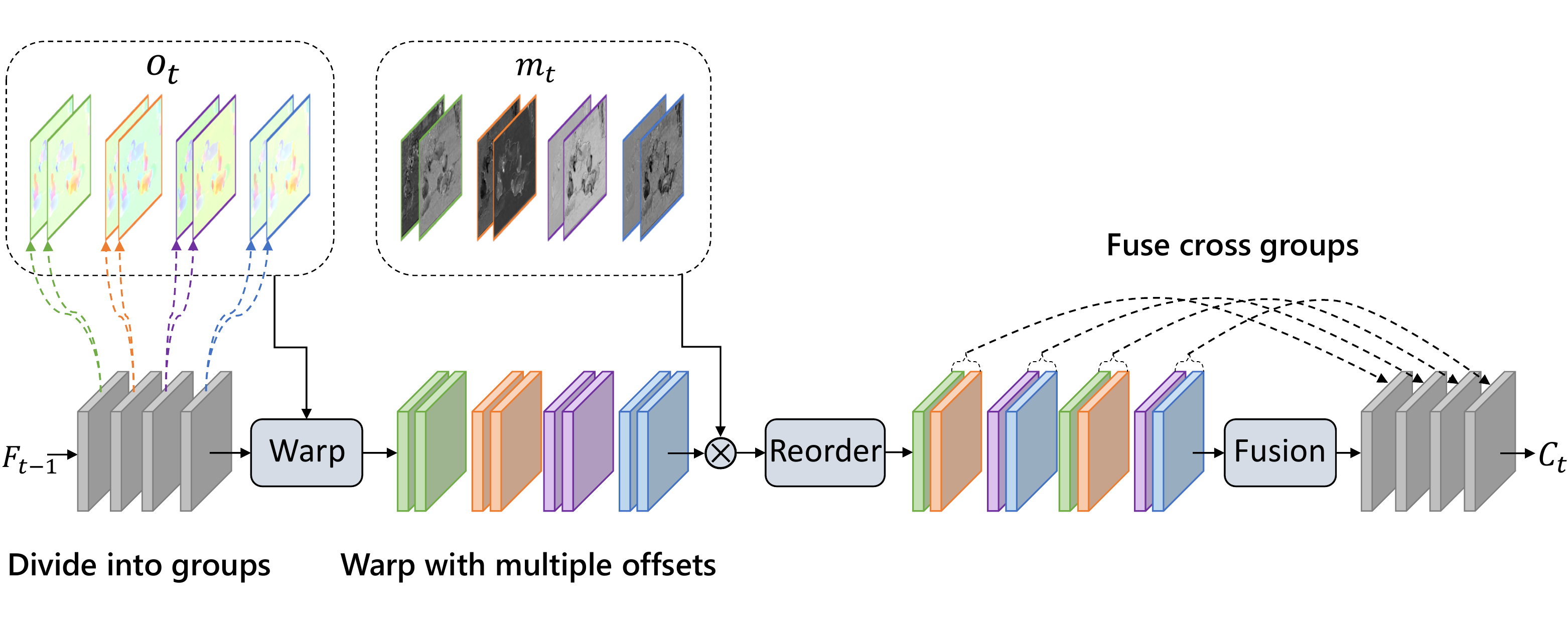}
		\end{center}
		\vspace{-0.5cm}
		\caption{Cross-group fusion module. In this example, the group number $G$ is 4 and the offset number $N$ of each group is 2.	}
		\vspace{-6mm}
		\label{OffsetDiversity}
	\end{figure}
	
In addition, motivated by the channel shuffle operation \cite{zhang2018shufflenet} which  improves the information flow in the CNN backbone, we customize a group-level interaction mechanism to further tap the potential of offset diversity for NVC.  
In particular, after warping each group with multiple offsets and applying the corresponding masks, we will reorder all groups before the fusion, as shown in Fig. \ref{OffsetDiversity}.  If using $g_{i}^{j}$ to represent the $i$-th group warped with its $j$-th offset, the features before reordering are $g_{0}^{0},...,g_{0}^{N-1},g_{1}^{0},...,g_{1}^{N-1},
\cdots,g_{G-1}^{0},...,g_{G-1}^{N-1}$, where the offset order is primary and group order is secondary. Then we reorder them as  $g_{0}^{0},...,g_{G-1}^{0},g_{1}^{1},...,g_{G-1}^{1},
\cdots,g_{0}^{N-1},...,g_{G-1}^{N-1}$, where the group order is primary instead. The following fusion operation will fuse every $N$ contiguous groups into one group. Therefore, during this process, the group reordering enables more cross-group interactions without increasing complexity. This design also enjoys the similar benefit with the weighted prediction from different reference frames in traditional codec.
Via the cross-group fusion, more diverse combinations in extracting temporal contexts from different groups are introduced and further improve the effectiveness of offset diversity. 
\label{offset_diversity}

\subsection{Quadtree Partition-Based Entropy Coding}
\label{quadtree}

\begin{figure}[t]
		\begin{center}
			\includegraphics[width=1\linewidth]{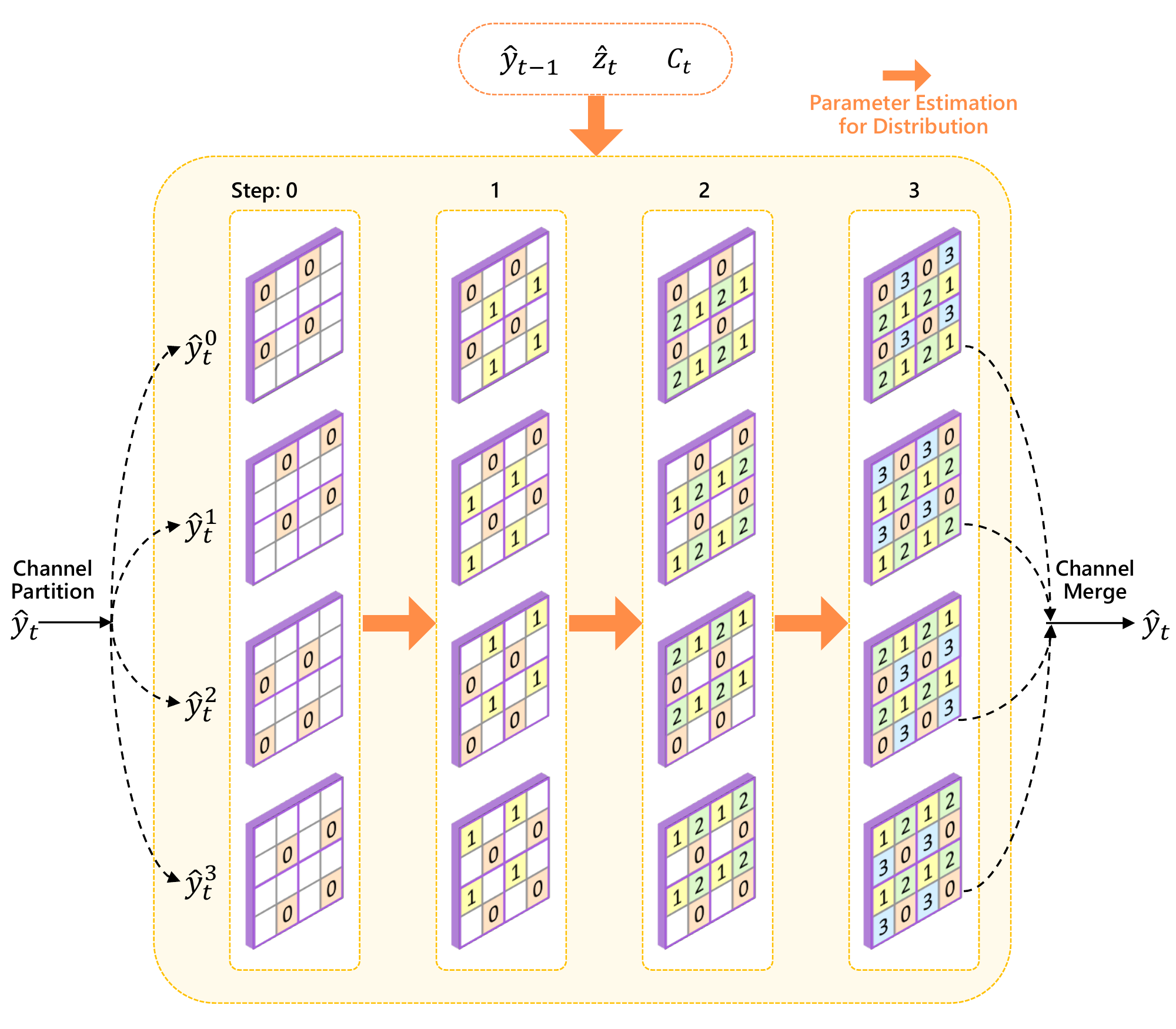}
		\end{center}
		\vspace{-0.6cm}
		\caption{Entropy coding with quadtree partition. The number means the coding order index. During the 4 coding steps, the  $\hat{y}_{t-1}$ from previous frame, hyper prior $\hat{z}_t$, and temporal context $C_t$ are also used for entropy modelling.	}
		\vspace{-5mm}
		\label{Quadtree_partition}
	\end{figure}
	
After obtaining temporal context $C_t$ via our offset diversity module, the input frame $x_t$ will be encoded and quantized as  $\hat{y}_t$, conditioned on $C_t$,   as shown in Fig. \ref{framework}.  We need to estimate the probability mass function (PMF) of $\hat{y}_t$ for its arithmetic coding. In this process, how to build an accurate entropy model to  estimate the PMF of  $\hat{y}_t$ is  vital for the compression efficiency.

Many neural codecs adopt the auto-regressive model \cite{minnen2018joint} as entropy model. However, it seriously slows down the coding speed. By contrast, the checkerboard model \cite{he2021checkerboard}  proposes coding the even positions of $\hat{y}_t$ first, and then use them to predict the PMF of the odd positions in parallel. Recently the dual spatial model \cite{li2022hybrid} improves it by utilizing the correlation along channel dimension. However, the neighbours used for entropy modelling in \cite{he2021checkerboard,li2022hybrid} are still limited when compared with  auto-regressive model. Thus, inspired from \cite{reed2017parallel, nakanishi2019neural}, this paper proposes a finer-grained coding manner via the quadtree partition, where  diverse spatial contexts are exploited to improve entropy modelling.

As shown in Fig. \ref{Quadtree_partition}, we first divide   $\hat{y}_t$	into four groups along the channel dimension. Then each group is partitioned into non-overlapped 2$\times$2 patches in spatial dimension. The whole entropy coding is divided into four steps, and each step codes the different positions associated with the corresponding indexes in Fig. \ref{Quadtree_partition}. At 0th step, all positions with index 0 of all patches are coded at the same time. It is noted, for the four groups, the positions with index 0 are different 
from each other. Thus, for every spatial position, there are one fourth channels (i.e., one group)  encoded. In the subsequent 1st, 2nd, and 3rd steps, all  positions coded in previous steps are used for predicting the PMF of the positions coded in the current step, and different spatial positions are coded for different groups in each step. 

During this process, more diverse neighbours are utilized. If considering the 8 spatial neighbours for a position, the auto-regressive model \cite{minnen2018joint} uses 4 (left,  top-left, top, top-right) neighbours for every position if not considering the boundary region. The checkerboard and dual spatial models \cite{he2021checkerboard,li2022hybrid} uses 0 and 4 (left, top, right, bottom) neighbours for the 0th and 1st steps, respectively. By contrast, as shown in Fig. \ref{Quadtree_partition}, our DCVC-DC uses 0, 4, 4, and 8 neighbours for the four steps, respectively. On average, the neighbour number in  DCVC-DC   is 2 times of that of  \cite{he2021checkerboard,li2022hybrid} and is same  with that of auto-regressive model. However, our model is much more time-efficient than auto-regressive model as all positions in each step can be coded in parallel. In addition,  our model also exploits the cross-channel correlation, which is like \cite{li2022hybrid} but in a refined way. For example, at the 3rd step, for one specific position of a group, the other channels at the same position are already coded  from   different groups in previous steps, and they can be used as the contexts for the entropy modelling in this step. This helps further squeeze the redundancy. Overall, our quadtree partition-based solution makes the entropy coding benefit from the finer-grained and diverse contexts, which fully mines the correlation from both spatial and channel dimensions.

\subsection{Implementation}
Our DCVC-DC is based on DCVC-HEM\cite{li2022hybrid} but focuses on exploiting Diverse Contexts to further boost performance. 
In addition, we also make the following improvements to obtain better tradeoff between performance and complexity. The first is that, considering depthwise separable convolution \cite{chollet2017xception} can reduce the computation cost while alleviating over-fitting,  we widely use it to replace the normal convolution in the basic block design. The second is that we use the unequal channel number settings for features with different resolutions, where the higher resolution feature is assigned with smaller channel number for acceleration. The third is that we  move partial quantization operations to higher resolution in the encoder, which helps achieve more precise bit allocation. The harmonization of encoding and quantization also brings some compression ratio improvements.
The section \ref{ablation} verifies the effectiveness of these structure optimizations, and the detailed network structures can be found in supplementary materials.

In particular, our DCVC-DC also already outperforms ECM in YUV420 colorspace. Not like traditional codec needs many hand-crafted changes in designing coding tools for different colorspaces, DCVC-DC only just needs simple adaptions based on a existing model trained for RGB. Without changing the network structure, we only up-sample the UV to use the unified input interface with RGB. Correspondingly, after obtaining the reconstructed frame, it is down-sampled on UV. In addition, our model for YUV420 just needs a simple finetune training based on the model trained for RGB.

\section{Experimental Results}

\begin{figure*}[t]
	\centering
	\includegraphics[width=1.0\linewidth]{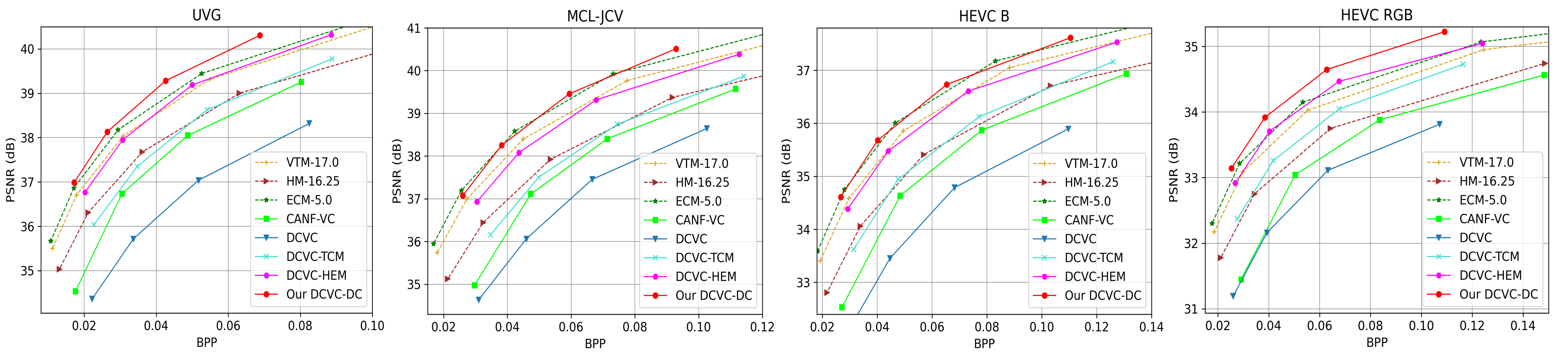}
	\vspace{-0.6cm}
	\caption{Rate and distortion curves. The comparison is in RGB colorspace measured with PSNR.  More results including  the corresponding MS-SSIM curves and the comparison in YUV420 colorsapce are in supplementary materials.
	}
		\vspace{-0.2cm}
	\label{RD_curve}
\end{figure*}

\begin{table*}[t]
  \centering
  \caption{BD-Rate (\%) comparison in RGB colorspace measured with PSNR. The anchor is VTM-17.0.}
   \renewcommand{\arraystretch}{1.2}
    \small
    \begin{tabular}{ccccccccc}
    \toprule[1.0pt]
                                         & UVG    & MCL-JCV  & HEVC B & HEVC C & HEVC D & HEVC E & HEVC RGB  & Average \\ \hline
VTM-17.0		                         & 0.0      & 0.0    & 0.0       & 0.0    &  0.0    & 0.0    & 0.0      & 0.0     \\ \hline
HM-16.25		                         & 36.4	    & 41.5	 & 38.8	     & 36.0	  &  33.7	& 44.0	 & 39.4		& 38.5     \\ \hline
ECM-5.0		                             & --10.0	& --12.2 & --11.5    & --13.4 &  --13.5	& --10.9 & --11.1	& -11.8     \\ \hline
CANF-VC	\cite{canfvc}	                 & 73.0	    & 70.8	 & 64.4	     & 76.2	  &  63.1	& 118.0	 & 79.9		& 77.9     \\ \hline
DCVC \cite{li2021deep}		             & 166.1    & 121.6	 & 123.2	 & 143.2  &  98.0	& 266.1	 & 113.4	& 147.4   \\ \hline
DCVC-TCM \cite{sheng2021temporal}		 & 44.1	    & 51.0	 & 40.2	     & 66.3	  &  37.0	& 82.7	 & 24.4		& 49.4     \\ \hline
DCVC-HEM \cite{li2022hybrid}		     & 1.1	    & 8.6	 & 5.1	     & 22.2	  &  2.4	& 20.5	 & -9.9		& 7.1     \\ \hline
Our DCVC-DC		                         & --19.1	& --11.3 & --12.0    & --10.3 &  --26.1	& --18.0 & --27.6	& --17.8     \\ 
    \bottomrule[1.0pt]
    \end{tabular}
    	\vspace{-0.5cm}
  \label{tab_rgb_psnr}
\end{table*}

\subsection{Experimental Settings}
\textbf{Datasets.} For training, we follow most existing NVCs and use Vimeo-90k \cite{xue2019video}. For testing YUV420 videos, HEVC B$\sim$E \cite{bossen2013common}, UVG \cite{mercat2020uvg}, and MCL-JCV \cite{wang2016mcl} are used. Their raw format is YUV420, so there is no any change before feeding them to NVC. For testing RGB videos, as these testsets have no RGB format, most existing NVCs use BT.601 (default in FFmpeg) to convert them from YUV420 to RGB. Actually,
JPEG AI \cite{jpeg_ai,jpeg_ai2} adopts BT.709  because
using BT.709 obtains higher compression ratio under similar visual quality. 
Thus, 
this paper follows JPEG AI and uses BT.709 for all codecs during testing RGB. 
It is noted that the relative bitrate comparisons between different codecs are similar in BT.601 and BT.709. 
The supplementary materials show the results using BT.601. In addition, we follow \cite{sheng2021temporal,li2022hybrid} and also test HEVC RGB dataset \cite{flynn16common} when testing RGB videos, and there is no format change as HEVC RGB dataset itself is in RGB format.

\textbf{Test Conditions.} We follow \cite{sheng2021temporal,li2022hybrid}  and test 96 frames for each video, and the intra period is set as 32. The low delay encoding setting is used, as the same with most existing works \cite{lu2019dvc,agustsson2020scale, li2021deep}. 
BD-Rate \cite{bjontegaard2001calculation} is used to measure the compression ratio, where negative numbers indicate bitrate saving and positive numbers indicate bitrate increase.

Our benchmarks include  HM \cite{HM} and VTM \cite{VTM} which represent the best H.265 and H.266 encoder, respectively. In particular, we also compare with ECM \cite{ECM} which is the prototype of next generation traditional codec. For the codec setting, we follow \cite{sheng2021temporal,li2022hybrid} and further use 10-bit as the intermediate representation  when testing RGB, which leads to better compression ratio for the three traditional codecs. The detailed settings are shown in supplementary materials. As for the NVC benchmarks, we compare with the recent SOTA models including CANF-VC \cite{canfvc}, DCVC \cite{li2021deep}, DCVC-TCM \cite{sheng2021temporal}, and DCVC-HEM \cite{li2022hybrid}.

\textbf{Model Training.}  We adopt the multi-stage training strategy as \cite{sheng2021temporal, li2022hybrid}. Our model also supports variable bitrate in single model \cite{li2022hybrid}, so different $\lambda$ values are used in different optimization steps. We follow \cite{li2022hybrid} and use 4 $\lambda$ values (85, 170, 380, 840). 
But different from \cite{li2022hybrid} using constant distortion weight in the loss, this paper propose using hierarchical weight setting on  $w_t$ for the distortion term (the whole loss is defined in Fig. \ref{framework}). Considering our training set Vimeo-90k only has 7 frames for each video, we refer  traditional codec setting and set the pattern size  as 4. The $w_t$ settings for 4 consecutive frames are  (0.5, 1.2, 0.5, 0.9).

\subsection{Comparisons with Previous SOTA Methods}

\begin{table*}[t]
  \centering
  \caption{BD-Rate (\%) comparison in RGB colorspace measured with MS-SSIM. The anchor is VTM-17.0.}
   \renewcommand{\arraystretch}{1.2}
    \small
    \begin{tabular}{ccccccccc}
    \toprule[1.0pt]
                                         & UVG    & MCL-JCV  & HEVC B & HEVC C & HEVC D & HEVC E & HEVC RGB  & Average    \\ \hline
VTM-17.0		                        & 0.0      & 0.0    & 0.0       & 0.0    &  0.0    & 0.0    & 0.0      & 0.0     \\ \hline
HM-16.25	                            &31.1	  &38.8	    &36.6	 &35.2	    &33.0	  &41.1	    &36.6		&36.1     \\ \hline
ECM-5.0	                                &--9.1	  &--11.1	&--10.2	 &--11.7	&--11.0	  &--9.9	&--9.8		&--10.4    \\ \hline
CANF-VC \cite{canfvc}	                &46.5	  &26.0	    &43.5	 &30.9	    &17.9	  &173.0	&57.7		&56.5     \\ \hline
DCVC \cite{li2021deep}		            & 64.9    & 27.5    & 54.4	 & 39.7     &  15.2	  & 210.4	& 51.3	    & 66.2   \\ \hline
DCVC-TCM \cite{sheng2021temporal}		&1.0      &--10.8	&--11.7	 &--15.2	&--29.0	  &16.7	    &--22.2		&--10.2   \\ \hline
DCVC-HEM \cite{li2022hybrid}	        &--25.2	  &--36.3	&--38.0	 &--38.3	&--48.1	  &--25.8	&--43.6		&--36.5  \\ \hline
Our DCVC-DC	                            &--32.6	  &--44.8	&--47.8	 &--49.8	&--58.2	  &--45.8	&--54.4		&--47.6     \\ 

    \bottomrule[1.0pt]
    \end{tabular}
  \label{tab_rgb_msssim}
\end{table*}

\begin{table*}[t]
  \centering
  \caption{BD-Rate (\%) comparison in YUV420 colorspace measured with PSNR. The anchor is VTM-17.0.}
   \renewcommand{\arraystretch}{1.2}
    \small
    \begin{tabular}{ccccccccc}
    \toprule[1.0pt]
                                         & UVG    & MCL-JCV  & HEVC B & HEVC C & HEVC D   & HEVC E        & Average        \\ \hline

							VTM-17.0	& 0.0      & 0.0    & 0.0       & 0.0    &  0.0     & 0.0       & 0.0            \\ \hline
							HM-16.25	&36.7	   &42.5	&39.2	    &33.3	 &  30.0	&40.7		&37.1            \\ \hline
							ECM-5.0	    &--10.6	   &--13.7	&--12.6	    &--14.7	 &--14.9	&--12.1		&--13.1          \\ \hline
							Our DCVC-DC &--17.8	   &--12.0	&--10.8	    &--12.4	 &--28.5	&--20.4		&--17.0          \\

    \bottomrule[1.0pt]
    \end{tabular}
    	\vspace{-0.3cm}
  \label{tab_yuv_psnr}
\end{table*}

\textbf{RGB colorspace.} Table \ref{tab_rgb_psnr} and \ref{tab_rgb_msssim} show the BR-rate comparison using RGB videos in terms of PSNR and MS-SSIM, respectively.  From Table \ref{tab_rgb_psnr}, we find our codec achieves significant compression ratio improvement over VTM on every dataset, and there is an average of 17.8\% bitrate saving. By contrast, the other neural codecs are still worse than  VTM. If using DCVC-HEM as anchor, our average bitrate saving is 23.5\%. In addition, our DCVC-DC also outperform ECM from Table \ref{tab_rgb_psnr}. If using ECM as anchor, an average of 6.4\% bitrate saving is achieved. 

Fig. \ref{RD_curve} shows the rate-distortion curves. From the curves, we can see our  DCVC-DC  achieves the SOTA compression ratio in wide bitrate range.
When using MS-SSIM as quality metric, our  DCVC-DC shows larger improvement. As shown in Table \ref{tab_rgb_msssim},  DCVC-DC has an average of 47.6\% bitrate saving over VTM. By contrast, the corresponding number of ECM over VTM is only 10.4\%.

It is noted that Table \ref{tab_rgb_psnr} and \ref{tab_rgb_msssim} use the RGB video with BT.709 conversion.
If with BT.601 conversion, the relative bitrate saving is similar with that in BT.709. For example,  with BT.601 conversion,  DCVC-DC over VTM  has an average of 18.0\% bitrate saving in terms of PSNR. More results with BT.601 can be seen in supplementary materials.

\textbf{YUV420 colorspace.} Actually traditional codecs are mainly optimized in YUV420. Thus, the comparison in YUV420 is also very important for evaluating the progress of NVC over traditional codec. The corresponding results are shown in Table \ref{tab_yuv_psnr}. The numbers in this table are calculated using the weighted PSNR for the three components of YUV. The weights are (6,1,1)/8, which are consisted with that in standard committee \cite{sullivan2011meeting}. 
As most NVCs have no corresponding released models for YUV420, Table \ref{tab_yuv_psnr} only reports the numbers of our NVC. We can see that  DCVC-DC has an average of 17.0\% bitrate saving over VTM. If only considering the Y component, an average   15.3\% bitrate saving is achieved over VTM.
Better yet, our  DCVC-DC also outperforms ECM in YUV420 on average, as shown in Table \ref{tab_yuv_psnr}.
This is an important milestone in the development of NVC. It is noted  our codec uses the same network structure for both RGB and YUV420 colorspaces, where only different finetunings are used during training. This shows the simplicity and strong extensibility of our  codec on the optimization for different input colorspaces. 

\subsection{Ablation Study}
\label{ablation}

\begin{table}[t]
\caption{Ablation Study on Diverse Contexts.}
\centering
\scalebox{0.9}{
\renewcommand{\arraystretch}{0.86}
\begin{tabular}{cccccc}
\toprule[1.0pt]
          
                                & \Gape[0.18cm][0.18cm]{ $M_a$}    & $M_b$                      & $M_c$                    & $M_d$                   & $M_e$                     \\ \hline
   Hierarchical quality         &   \multirow{2}*{\checkmark}    &                            &                          &                         &                           \\     
          structure             &                                &                            &                          &                         &                           \\ \hline    
   Offset diversity             &   \multirow{2}*{\checkmark}    &\multirow{2}*{\checkmark}   &                          &                         &                           \\     
    w/ cross-group              &                                &                            &                          &                         &                           \\ \hline    
   Offset diversity             &                                &                            &\multirow{2}*{\checkmark} &                         &                           \\     
w/o cross-group \cite{ODPaper}  &                                &                            &                          &                         &                           \\ \hline    
   Quadtree partition           &   \multirow{2}*{\checkmark}    &\multirow{2}*{\checkmark}   &\multirow{2}*{\checkmark} &\multirow{2}*{\checkmark}&                           \\     
     based model                &                                &                            &                          &                         &                           \\ \hline    
   Dual spatial                 &                                &                            &                          &                         & \multirow{2}*{\checkmark} \\     
model\cite{li2022hybrid}        &                                &                            &                          &                         &                           \\ \hline    
\Gape[0.2cm][0.2cm]{ BD-Rate(\%)}   &            0.0                 &         8.4                &        12.1              &       14.7              &  21.3                         \\     
\bottomrule[1.0pt] 
\end{tabular}
}
	\vspace{-5mm}
\label{tab_abalation}
\end{table}
 
To verify the effectiveness of each component, we conduct  comprehensive ablation studies. For simplification, the HEVC datasets in RGB colorspace are used here. The average BD-Rate in terms of PSNR is shown.

\textbf{Diverse Contexts.} Table \ref{tab_abalation} shows the study on the effectiveness of diverse contexts. 
First, from the comparison between $M_e$ and $M_d$, we can see that the BD-rate is reduced from 21.3\%  to 14.7\%. This large difference shows the substantial coding gain of our qaudtree partition-based entropy coding, and verifies the advantages of diverse spatial and channel contexts via finer-grained partition.  

From temporal dimension, we also  design hierarchical quality structure and offset diversity with cross-group interaction. In Table \ref{tab_abalation}, based on $M_d$, we first test the original   offset diversity \cite{ODPaper} without cross-group interaction, i.e., removing the reorder operation in Fig. \ref{OffsetDiversity}. However, it ($M_c$) only brings 2.6\% BD-rate difference. By contrast, powered by our cross-group interaction, the potential of offset diversion is fully tapped, and $M_b$ reduces the BD-rate number by 6.3\% over $M_d$. At last, based on $M_b$, we evaluate the  hierarchical quality structure, i.e., $M_a$. The  8.4\% gap shows   learning   high-quality contexts brings large benefits to the mining of temporal correlation across many frames.

\textbf{Structure optimization.} Although our codec learns utilizing diverse contexts in efficient manner, we still purse better tradeoff between compression ratio and computational cost. Therefore, we further optimize our model in network structure. Table \ref{tab_optimization} shows the study. 
Based on the $M_a$ (same with that in Table \ref{tab_abalation}), we first implement the depthwise separable convolution into codec. $M_h$ shows  widely using depthwise separable convolution to replace the normal convolution not only significantly reduces the MACs, but also brings some compression ratio improvements. 

The second acceleration is that we use the unequal channel number settings. Not like many existing NVCs use the same channel number for features with different resolutions, we propose assigning the larger number for low resolution feature to increase the latent representation capacity while using the smaller number for  the high resolution feature 
to  accelerate model. The performance of $M_g$ verifies the effectiveness of our improvement. In addition, many existing NVCs perform the quantization at the low-resolution latent representation after encoding. To achieve more precise rate adjustment, this paper moves partial quantization operations to the higher resolution in the encoder. $M_f$ shows the integration of encoding and quantization brings some BD-rate improvements with negligible MAC change.

\begin{table}[t]
\caption{Ablation Study on Structure Optimization.}
\centering
\scalebox{0.85}{
\renewcommand{\arraystretch}{1.3}
\begin{tabular}{cccccc}
\toprule[1.0pt]
          
                                & $M_f$                          & $M_g$                      & $M_h$                        & $M_a$                  \\ \hline
Quant at high resolution        &     \checkmark                 &                            &                              &                        \\ \hline
Unequal channel setting         &     \checkmark                 &    \checkmark              &                              &                        \\ \hline    
  Depthwise separable conv              &     \checkmark                 &    \checkmark              &     \checkmark               &                        \\ \hline       
MACs                            &     2642G                      &      2642G                 &        2939G                 &      3456G             \\ \hline     
 BD-Rate (\%)                       &            0.0                 &         1.1                &        2.4                   &       3.5              \\     
\bottomrule[1.0pt]
\end{tabular}
}
	\vspace{-2mm}
\label{tab_optimization}
\end{table}

\subsection{Complexity}
The complexity comparison is shown in Table \ref{tab_complexity}. 
We find the MACs of our  DCVC-DC are reduced by  19.4\% when compared with  DCVC-HEM \cite{li2022hybrid}. However, the actual encoding and decoding time is higher. This is because currently the computational density of depthwise convolution is not as high as  normal convolution under the same MAC condition. But through the customized optimization \cite{lu2021optimizing}, it can be further accelerated in the future. From another perspective, considering that our  DCVC-DC has 23.5\% bitrate saving over previous SOTA DCVC-HEM \cite{li2022hybrid}, such increase degree in running time is a price worth paying. By contrast, ECM brings 13.1\% (Table \ref{tab_yuv_psnr}) improvement over its predecessor VTM, but the encoding complexity is more than 4 times \cite{ECM_performance} of VTM.

\begin{table}[t]
\caption{Complexity comparison.}
\centering
\scalebox{0.84}{
\renewcommand{\arraystretch}{1.2}
\begin{tabular}{ccccc}
\toprule[1.0pt]
          &           MACs    & Encoding   Time      & Decoding   Time               \\ \hline
  									
DCVC-HEM\cite{li2022hybrid}            & 3279G   & 890ms         & 652ms    \\ \hline
Our DCVC-DC                            & 2642G   & 1005ms         & 765ms    \\
\bottomrule[1.0pt]
\end{tabular}
}
\\
\vspace{0.2cm}
\raggedright\footnotesize{ \, Note:  Tested on NVIDIA 2080TI with using 1080p as input.}
   \vspace{-0.7cm}
\label{tab_complexity}
\end{table}

\section{Conclusion and Limitation}
In this paper, we have presented how to utilize diverse contexts to further boost NVC. From temporal dimension, the model is guided to extract long-term and yet high-quality contexts to alleviate error propagation and exploit long range correlation. The offset diversity with cross-group interaction provides complementary motion alignments to handle complex motion. From spatial dimension, the fine-grained quadtree-based partition is proposed to increase spatial context diversity. Powered by our techniques, the compression ratio of NVC has been pushed to new height. Our   DCVC-DC has surpassed ECM in both RGB and YUV420 colorspaces, which is an important milestone in the development of NVC.

During the training, to learn the hierarchical quality pattern, we still use the fixed distortion weights which are similar with those in traditional codec. This  may not be the best choice for NVC. Actually, reinforcement learning is  good at solving such kind of time series weight decision problem. In the future, we will investigate utilizing reinforcement learning to help NVC make better weight decision with considering the temporal dependency.

{\small

}

\clearpage
\begin{appendices}

Appendices provide  the supplementary material to our proposed neural video compression (NVC) with diverse contexts, i.e., DCVC-DC model.

\section{Network Structure}
Our DCVC-DC is based on DCVC-HEM \cite{li2022hybrid}, but focuses more on exploiting Diverse Contexts to further boost compression  efficiency. The learning of hierarchical quality pattern  is mainly performed in training phase via adjusting the distortion weight in the loss. Here we describe other implementation details in  network structure.

\begin{figure}[t]
	\begin{center}
		\includegraphics[width=0.8\linewidth]{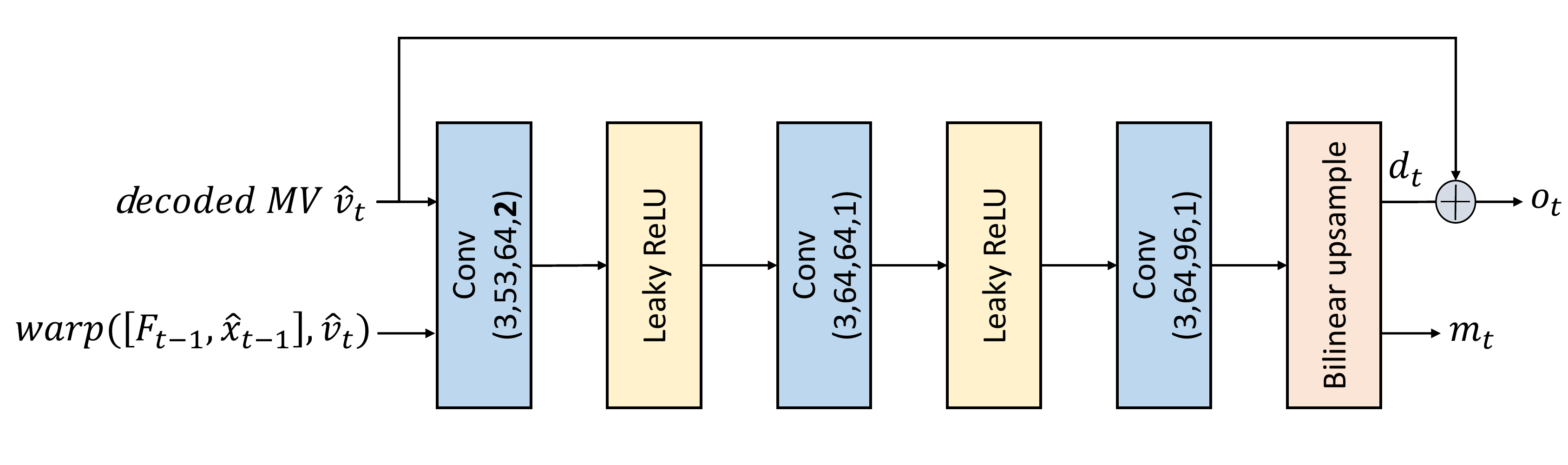}
	\end{center}
	\vspace{-0.5cm}
	\caption{The network structure of offset prediction.}
	\vspace{-3mm}
	\label{supp_offset_prediction}
\end{figure}

\textbf{Group-based offset diversity.} In this process, we will predict a total of $G \times N$ offsets $d_t$, where $G$ is the group number and $N$ is offset number for each group. In the implementation, the channel dimension of propagated feature $F_{t-1}$ is 48. It is divided into 16 groups and each group has 2  offsets, i.e., $G$=16 and $N$=2. The detailed network structure of offset  prediction is shown in Fig. \ref{supp_offset_prediction}. 
For convolution layer, the ($K$, $Cin$, $Cout$, $S$) indicate the kernel
size, input channel number, output channel number, and stride, respectively
The inputs include the decoded motion vector (MV) $\hat{v}_t$. In addition, the previous reconstructed frame $\hat{x}_t$ and propagated $F_{t-1}$ are also warped
and fed as the auxiliary information. The outputs include the residual offsets $d_t$. $d_t$ adds the  $\hat{v}_t$ to get the final offsets $o_t$. In addition, the corresponding masks $m_t$ are also generated.
It is noted that, the first convolution layer will reduce the resolution by 2x for acceleration. After the last convolution layer, we use  bilinear to upsample them back to original resolution.  

\begin{figure}[t]
	\begin{center}
		\includegraphics[width=1\linewidth]{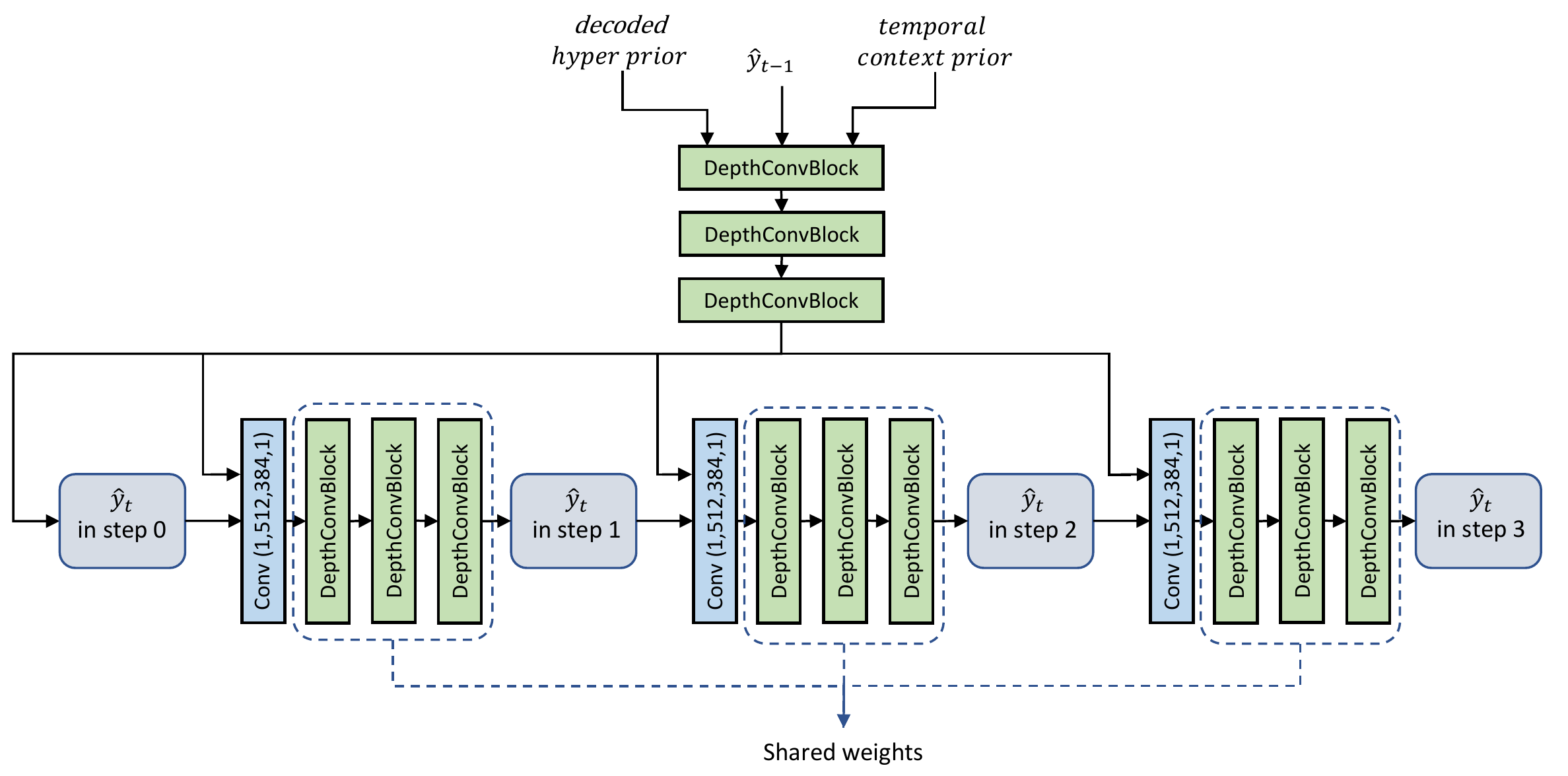}
	\end{center}
	\vspace{-0.5cm}
	\caption{The network structure of entropy model.}
	\vspace{-3mm}
	\label{supp_quadtree_partition}
\end{figure}

\textbf{Quadtree partition-based entropy coding.} The proposed entropy coding can be classified into 4 steps. The network structure is shown in Fig. \ref{supp_quadtree_partition}. As shown in this figure, the quantized latent representation $\hat{y}_{t-1}$ from the previous frame, hyper prior $\hat{z}_t$, and temporal context $C_t$ are also used for predicting the distribution parameters for all $\hat{y}_{t}$ in the 4 steps. In addition, all positions coded in previous steps will also be used for predicting the  distribution parameters of the positions coded in the current steps. In this process, to reduce the model parameters, most network blocks therein use the shared weights, as shown in Fig. \ref{supp_quadtree_partition}.

\begin{figure}[t]
	\begin{center}
		\includegraphics[width=0.75\linewidth]{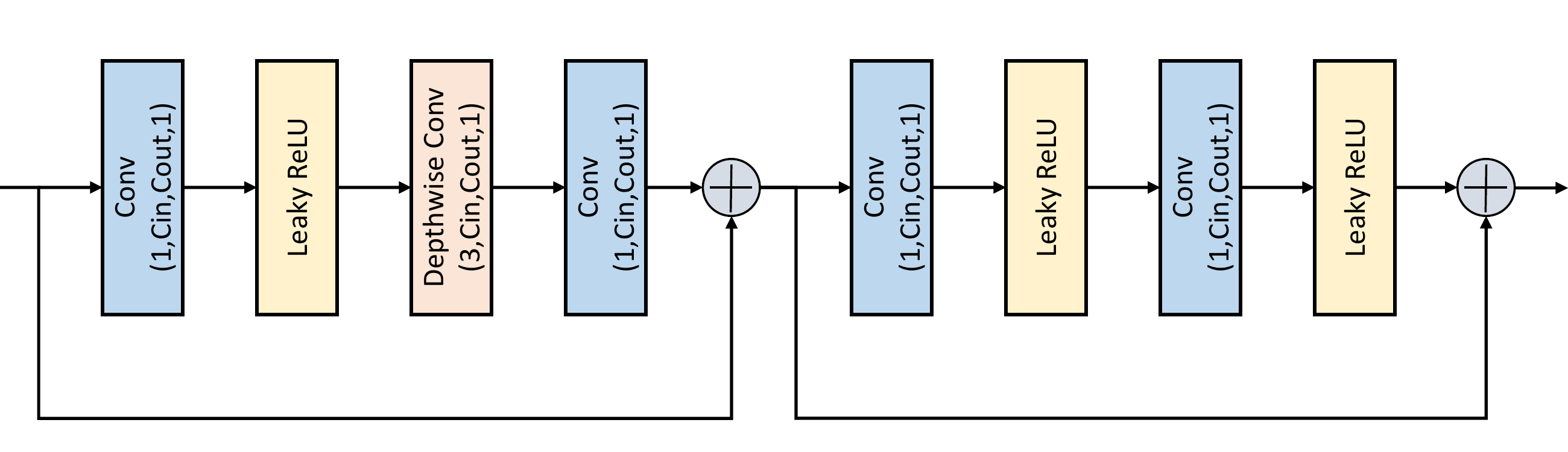}
	\end{center}
	\vspace{-0.5cm}
	\caption{The network structure of DepthConvBlock. }
	\vspace{-3mm}
	\label{supp_depthConvBlock}
\end{figure}

\textbf{Structure optimization.} As mentioned in the main paper, to reduce the computation cost, we widely adopt the depthwise separable convolution. As shown in Fig. \ref{supp_quadtree_partition}, the basic block in our entropy model  is DepthConvBlock which contains depthwise separable convolution. The structure of  DepthConvBlock  is shown Fig. \ref{supp_depthConvBlock}. In the DepthConvBlock, except that the depthwise convolution layer is with 3x3 kernel, all regular convolution layers use 1x1 kernel to further reduce the computation cost.

\begin{figure}[t]
	\begin{center}
		\includegraphics[width=1\linewidth]{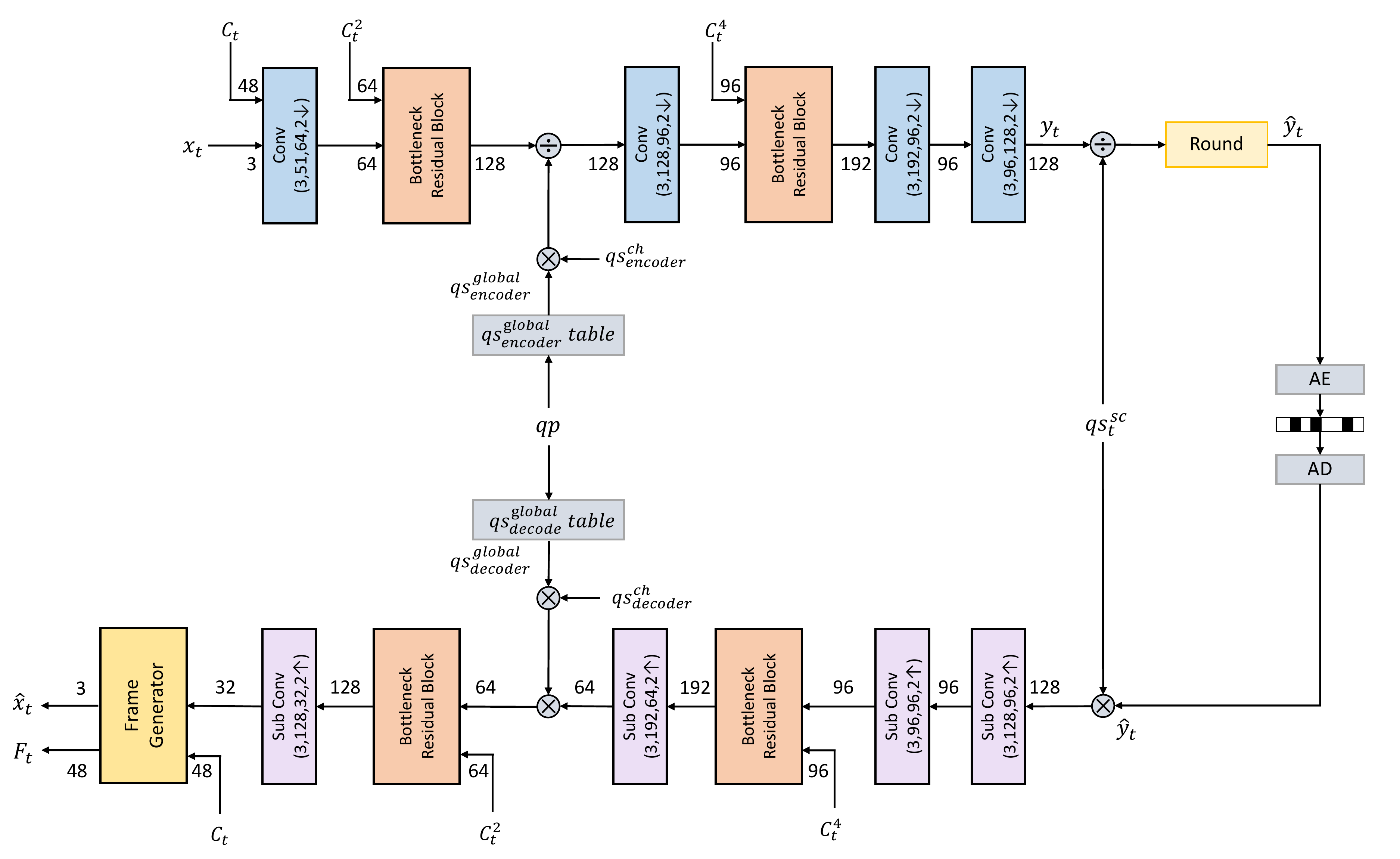}
	\end{center}
	\vspace{-0.5cm}
	\caption{The network structure of contextual encoder and decoder. The frame generator follows the decoder. }
	\vspace{-3mm}
	\label{supp_contextual_encdec}
\end{figure}

In addition, we use the unequal channel number settings for the encoder and decoder. The network structure of our contextual encoder and decoder for frame coding is shown in Fig. \ref{supp_contextual_encdec}. From this figure, we can see that the propagated feature  $F_t$ and the motion-aligned $C_t$ with high resolution are with 48 channel number. They are smaller than 64 used in \cite{li2022hybrid}. It can help us  reduce the computation cost. At the same time, the quantized latent representation $\hat{y}_t$ uses 128 channel number, and it is larger than 96 used in \cite{li2022hybrid}. This can bring some compression ratio improvements as  the latent representation has larger capacity. By adjusting the channel number for features with different resolutions, a better trade-off between compression ratio and computation cost can be achieved.  
In Fig. \ref{supp_contextual_encdec}, we follow \cite{sheng2021temporal,li2022hybrid} and also adopt the multi-scale contexts $C_{t}^{2}$ and $C_{t}^{4}$, which are 2x and 4x down-sampled temporal contexts. Their generation details can be found in \cite{sheng2021temporal,li2022hybrid}. The bottleneck residual block and frame generator in Fig. \ref{supp_contextual_encdec} are similar with those in \cite{li2022hybrid}.

\begin{figure}[t]
	\begin{center}
		\includegraphics[width=1\linewidth]{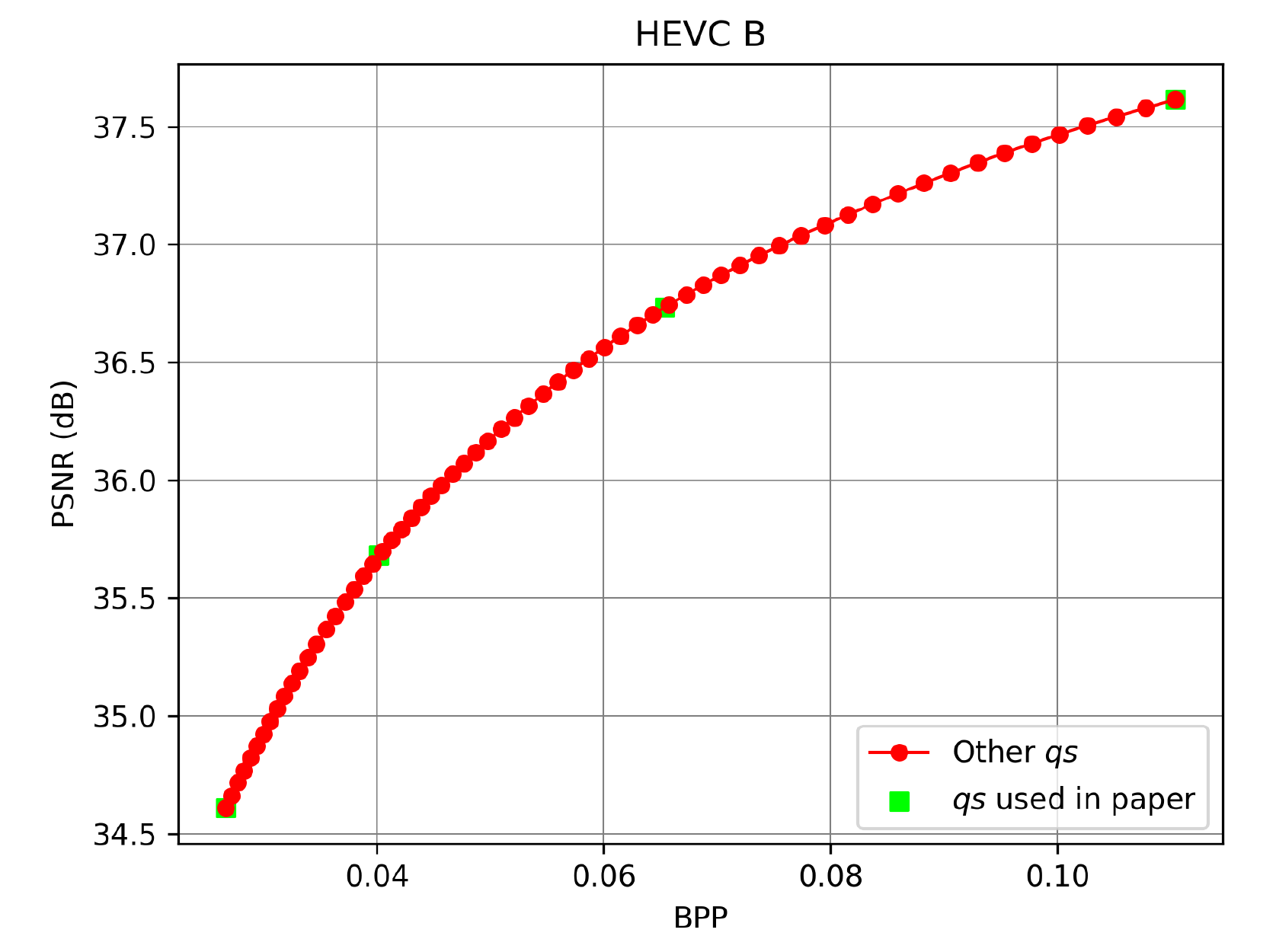}
	\end{center}
	\vspace{-0.7cm}
	\caption{Smooth rate adjustment in single model. }
	\vspace{-3mm}
	\label{supp_rate_adjustment}
\end{figure}

Our  codec supports variable bitrates in single model. For more precise rate adjustment, this paper proposes moving partial quantization operations to higher resolution. As shown in Fig. \ref{supp_contextual_encdec}, the quantization parameter $qp$ is used for controlling the bitrate via the user input. According to the $qp$, the global quantization step $qs_{encoder}^{global}$ is queried via the learnable quantization parameter-to-quantization step table. Then the learnable channel-wise $qs_{encoder}^{ch}$ is used to modulate the quantization step for each channel.  For $y_t$ at 16x down-sampled resolution, the spatial-channel-wise quantization step $qs_{t}^{sc}$ is applied before the rounding operation, where the $qs_{t}^{sc}$ for each frame is generated by the entropy model, like \cite{li2022hybrid}. During the decoding, the corresponding inverse operations are applied. This multi-granularity quantization mechanism originates from \cite{li2022hybrid}. However, in \cite{li2022hybrid}, global, 
channel-wise, and spatial-channel-wise quantization steps are all applied in the 16x down-sampled resolution. By contrast, we propose moving the global and 
channel-wise to the 2x down-sampled resolution for finer-grained adjustment. In addition, in our codec, the encoder and decoder have separate learnable global quantization step table and  channel-wise quantization step, which further enlarges the flexibility.   
As shown in Fig. \ref{supp_rate_adjustment}, we test 64 $qs$ values for our codec. From the curve, we can see that our codec  achieves very smooth rate adjustment in single model.

\section{Test Settings}

\begin{figure}[t]
	\begin{center}
		\includegraphics[width=1\linewidth]{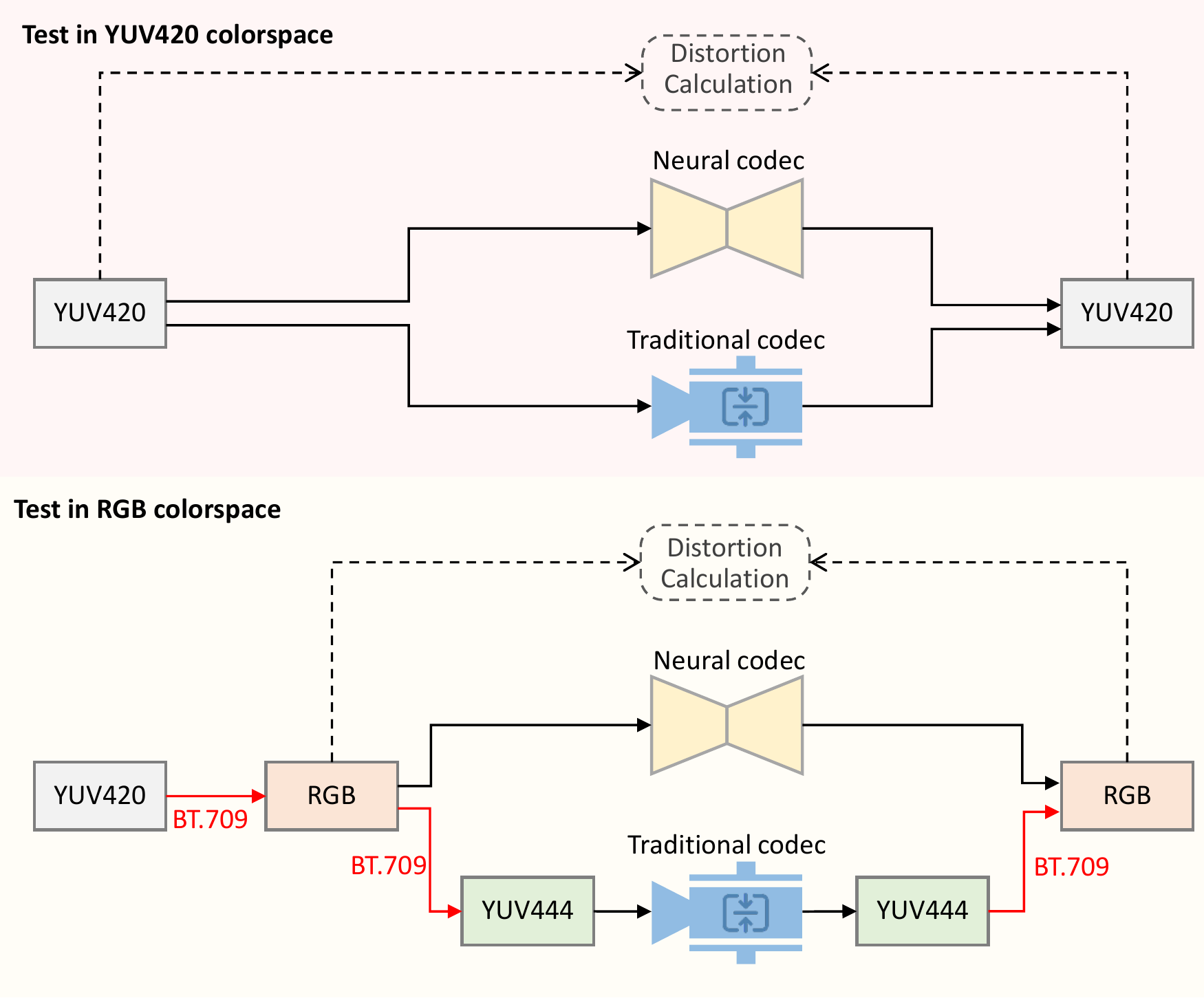}
	\end{center}
	\vspace{-0.5cm}
	\caption{Test pipeline in YUV420 and RGB colorspace, respectively. The red line indicates the colorspace conversion.  	}
	\vspace{-3mm}
	\label{supp_codec_space}
\end{figure}

To conduct  comprehensive comparisons, we compare the NVCs and traditional codecs in both YUV420 and RGB colorspaces. The test pipeline is shown in Fig. \ref{supp_codec_space}.

\textbf{YUV420.} When testing YUV420 video, there is no any colorspace conversion, as shown   in Fig. \ref{supp_codec_space}. For traditional codec, our benchmarks include HM \cite{HM}, VTM \cite{VTM}, and ECM \cite{ECM}. The three codecs use  \textit{encoder\_lowdelay\_main10.cfg}, \textit{encoder\_lowdelay\_vtm.cfg}, and  \textit{encoder\_lowdelay\_ecm.cfg} config files, respectively. The parameters for each video are as:
\begin{itemize}
	\item 
	-c \{{\em config file name}\}\par
	-\/-InputFile=\{{\em input video name}\}\par
	-\/-InputBitDepth=8\par
	-\/-OutputBitDepth=8 \par
	-\/-OutputBitDepthC=8 \par
	-\/-FrameRate=\{{\em frame rate}\}\par
	-\/-DecodingRefreshType=2\par
	-\/-FramesToBeEncoded=\{{\em frame number}\}\par
	-\/-SourceWidth=\{{\em width}\}\par
	-\/-SourceHeight=\{{\em height}\}\par
	-\/-IntraPeriod=32\par
	-\/-QP=\{{\em qp}\}\par
	-\/-Level=6.2\par
	-\/-BitstreamFile=\{{\em bitstream file name}\}\par
\end{itemize}

\begin{figure}[t]
	\begin{center}
		\includegraphics[width=1\linewidth]{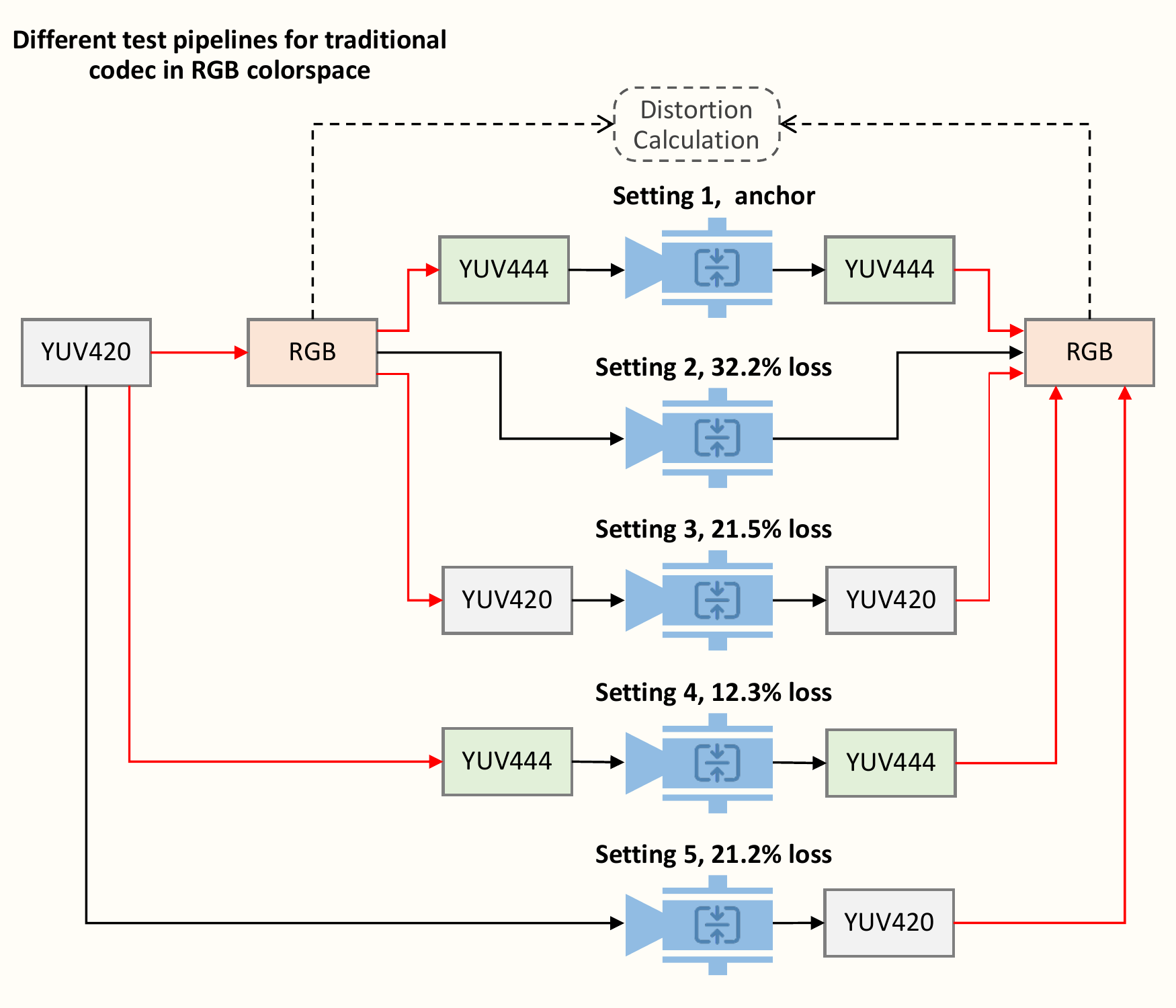}
	\end{center}
	\vspace{-0.5cm}
	\caption{Comparison of different input colorspaces for traditional codec when testing RGB videos. For the bitrate comparison, VTM-17.0  is used and tested on HEVC B dataset.	}
	\vspace{-3mm}
	\label{supp_codec_space_compare}
\end{figure}

\textbf{RGB.}  Except  that HEVC RGB testset is in RGB format, the raw formats of all other testsets are YUV420. Thus, to test RGB video, we need to convert them from YUV420 to RGB colorspace. Many existing NVC works use BT.601 (the default choice in FFmpeg) to conduct the conversion. Actually, JPEG AI  \cite{jpeg_ai,jpeg_ai2} uses BT.709 for the colorspace conversion. Thus, in the main paper, we follow JPEG  and also use BT.709 to convert the raw YUV420 video to RGB colorspace when testing RGB video. 
In addition, it is also noted that the raw YUV420 videos of HEVC datasets \cite{test_seq_format1,test_seq_format2,test_seq_format3} themselves are generated from RGB source using BT.709. However, it is unfortunate that currently we cannot access their raw RGB videos, and only have their YUV420 videos.
Therefore, when testing RGB video, we should use the same conversion manner (i.e., BT.709) to convert them back from YUV420 to RGB.

It is noted that, when traditional codecs test RGB videos, using YUV444 as the internal colorspace  achieves better compression ratio than directly using RGB, although the final distortion is measured in RGB.  Fig. \ref{supp_codec_space} shows that the RGB videos will be converted to YUV444 for higher compression ratio when testing traditional codecs. The reconstructed YUV444 videos will converted back to RGB for distortion calculation. 

Fig. \ref{supp_codec_space_compare} compares different test pipelines for traditional codec in testing RGB videos.  From this figure, we can see that using YUV444 as the internal colorspace (i.e., setting 1) achieves the best performance, and other settings have non-trivial bitrate increase.
Many existing works use setting 5 in Fig. \ref{supp_codec_space_compare}. However, we can see that there is 21.2\% bitrate increase when compared with the setting 1 we used.
Thus, to configure the best traditional codecs, we  use YUV444 as the internal colorspace.
For HM, VTM, and ECM,  \textit{encoder\_lowdelay\_main\_rext.cfg}, \textit{encoder\_lowdelay\_vtm.cfg}, and  \textit{encoder\_lowdelay\_ecm.cfg} config files are used, respectively.  The parameters for each video are as:
\begin{itemize}
	\item 
	-c \{{\em config file name}\}\par
	-\/-InputFile=\{{\em input file name}\}\par
	-\/-InputBitDepth=10\par
	-\/-OutputBitDepth=10 \par
	-\/-OutputBitDepthC=10 \par
	-\/-InputChromaFormat=444\par
	-\/-FrameRate=\{{\em frame rate}\}\par
	-\/-DecodingRefreshType=2\par
	-\/-FramesToBeEncoded=\{{\em frame number}\}\par
	-\/-SourceWidth=\{{\em width}\}\par
	-\/-SourceHeight=\{{\em height}\}\par
	-\/-IntraPeriod=32\par
	-\/-QP=\{{\em qp}\}\par
	-\/-Level=6.2\par
	-\/-BitstreamFile=\{{\em bitstream file name}\}\par
\end{itemize}

In addition, it is noted that ECM is still under development. As it is mainly optimized for YUV420, currently ECM-5.0 has several bugs on supporting YUV444 when  using it to test RGB videos.
We fixed them and verified the encoding and decoding match. After the bug fix, ECM-5.0 performs better than VTM-17.0, and the bitrate saving over VTM-17.0 is similar with that in YUV420. Thus, we believe the fix is reasonable for ECM-5.0 to support YUV444 coding.

\section{Results in RGB colorspace with BT.601}

Actually, when testing RGB videos, most existing NVC methods ignore the conversion manner and directly use BT.601 to conduct the conversion, because BT.601 is the default choice of FFmpeg. To make comparison with more existing NVCs, we also test our DCVC-DC under BT.601. It is noted that our DCVC-DC does not need any retraining for testing RGB videos with  BT.601. Table \ref{tab_601_res_psnr} and \ref{tab_601_res_msssim} show the BD-rate comparisons in terms of PSNR and MS-SSIM, respectively. 

In this two tables, we use a newer version of VTM, i.e., VTM-17.0 as the anchor, when compared with the VTM-13.2 used in \cite{li2022hybrid}. At the same time, we use the 10 bit   intermediate representation for YUV444 rather than 8 bit used in \cite{li2022hybrid}. These two modifications brings a more powerful baseline. As shown in Table \ref{tab_601_res_psnr}, the VTM-13.2 used in \cite{li2022hybrid} has an average 5.0\% bitrate increase than VTM-17.0 used in this paper. 

When using the stronger VTM-17.0 as anchor, we can see that our DCVC-DC also achieves significant bitrate saving  for RGB videos converted using BT.601. For example, Table \ref{tab_601_res_psnr} shows that our DCVC-DC can achieve an average of 18.0\% bitrate saving over VTM-17.0. By contrast, other NVCs still cannot surpass VTM-17.0.
These results verify the effectiveness of our DCVC-DC.

\begin{table*}[t]
	\centering
	\caption{BD-Rate (\%) comparison for RGB colorspace with BT.601. Quality is measured with PSNR. The anchor is VTM-17.0.}
	\renewcommand{\arraystretch}{1.25}
	\small
	\begin{tabular}{ccccccccc}
		\toprule[1.0pt]
		& UVG    & MCL-JCV  & HEVC B   & HEVC C     & HEVC D     & HEVC E    & HEVC RGB  & Average \\ \hline
		
		VTM-17.0                               & 0.0     & 0.0       &0.0        &  0.0        &  0.0      &   0.0     &   0.0     &    0.0 \\ \hline
		VTM-13.2   (from \cite{li2022hybrid} ) & 8.8     & 6.3       &3.2        &  1.9        &  0.5      &   8.6     &   5.5     &    5.0 \\ \hline
		HM-16.20   (from \cite{li2022hybrid} ) & 48.8    & 51.2      & 43.5      &  40.9       &  35.8     &   59.3    &   51.1    &    47.2 \\ \hline
		DVCPro \cite{lu2020end}                & 238.1   & 176.1     &  194.8    &   204.5     &   157.9   &   455.1   &  179.2    &    229.4 \\ \hline
		MLVC  \cite{lin2020m}                  & 137.6   & 140.0     &  126.0    &   216.7     &   165.7   &   262.2   &  163.5    &    173.1 \\ \hline
		RLVC \cite{yang2021learning}           & 244.0   & 221.3     &  205.1    &   202.7     &   141.8   &   398.2   &  199.4    &    230.4 \\ \hline
		CANF-VC \cite{canfvc}                  & 61.4    & 60.5      & 56.4      &  70.5       &  52.8     &   119.7   &  79.9     &    71.6 \\ \hline
		DCVC  \cite{li2021deep}                & 140.3   & 107.2     &  117.9    &   151.5     &   106.7   &   269.5   &  111.9    &    143.6 \\ \hline
		DCVC-TCM \cite{sheng2021temporal}      & 29.9    & 39.4      & 32.7      &  62.4       &  27.8     &   80.4    &  24.4     &    42.4 \\ \hline
		DCVC-HEM \cite{li2022hybrid}           &  --7.7  &   1.1     & --1.1     &   16.9      &  --8.4    &   20.8    &  --9.9    &    1.7   \\ \hline
		Our DCVC-DC                            & --21.0  &   --13.3  &  --13.7   &     --8.2   &   --27.9  &  --14.4   &  --27.6   &    --18.0  \\ 
		
		\bottomrule[1.0pt]
	\end{tabular}
	\label{tab_601_res_psnr}
\end{table*}

\begin{table*}[t]
	\centering
	\caption{BD-Rate (\%) comparison for RGB colorspace with BT.601. Quality is measured with MS-SSIM. The anchor is VTM-17.0.}
	\renewcommand{\arraystretch}{1.25}
	\small
	\begin{tabular}{ccccccccc}
		\toprule[1.0pt]
		& UVG    & MCL-JCV  & HEVC B   & HEVC C     & HEVC D     & HEVC E    & HEVC RGB  & Average \\ \hline
		
		VTM-17.0                              & 0.0     & 0.0      &0.0       & 0.0      & 0.0      & 0.0       &  0.0     &     0.0 \\ \hline
		VTM-13.2 (from \cite{li2022hybrid} )  & 3.4     & 4.6      &2.6       & 1.8      & 0.5      & 13.1      &  3.9     &    4.3 \\ \hline
		HM-16.20 (from \cite{li2022hybrid} )  & 37.2    & 46.3     &40.2      & 39.4     &  36.0    & 59.3      &  44.7    &    43.3 \\ \hline
		DVCPro \cite{lu2020end}               & 72.3    & 43.5     &64.5      & 61.6     &  24.3    & 248.1     &  67.8    &    83.2 \\ \hline
		RLVC \cite{yang2021learning}          & 86.2    & 77.6     &68.5      & 79.5     &  35.0    & 311.8     &  68.0    &    103.8 \\ \hline
		CANF-VC  \cite{canfvc}                & 31.2    & 14.2     &30.7      &  26.3    &   11.4   &  160.8    &   57.7   &     47.5 \\ \hline
		DCVC \cite{li2021deep}                & 37.1    & 10.2     &33.8      &  25.5    &   2.2    & 158.4     &  38.4    &    43.7 \\ \hline
		DCVC-TCM \cite{sheng2021temporal}     & --7.5   & --19.3   &  --21.5  &  --21.1  &  --36.2  &   12.6    & --22.2   &     --16.5 \\ \hline
		DCVC-HEM \cite{li2022hybrid}          & --32.6  &  --42.7  &   --45.7 &   --42.5 &   --54.5 &  --28.2   &  --43.6  &      --41.4 \\ \hline
		Our DCVC-DC                           & --37.5  &  --49.4  &   --53.4 &   --54.0 &   --63.1 &  --49.7   &  --54.4  &      --51.6 \\

		\bottomrule[1.0pt]
	\end{tabular}
	\label{tab_601_res_msssim}
\end{table*}

\section{Rate-Distortion Curves}
\begin{figure*}[t]
	\begin{center}
		\includegraphics[width=0.8\linewidth]{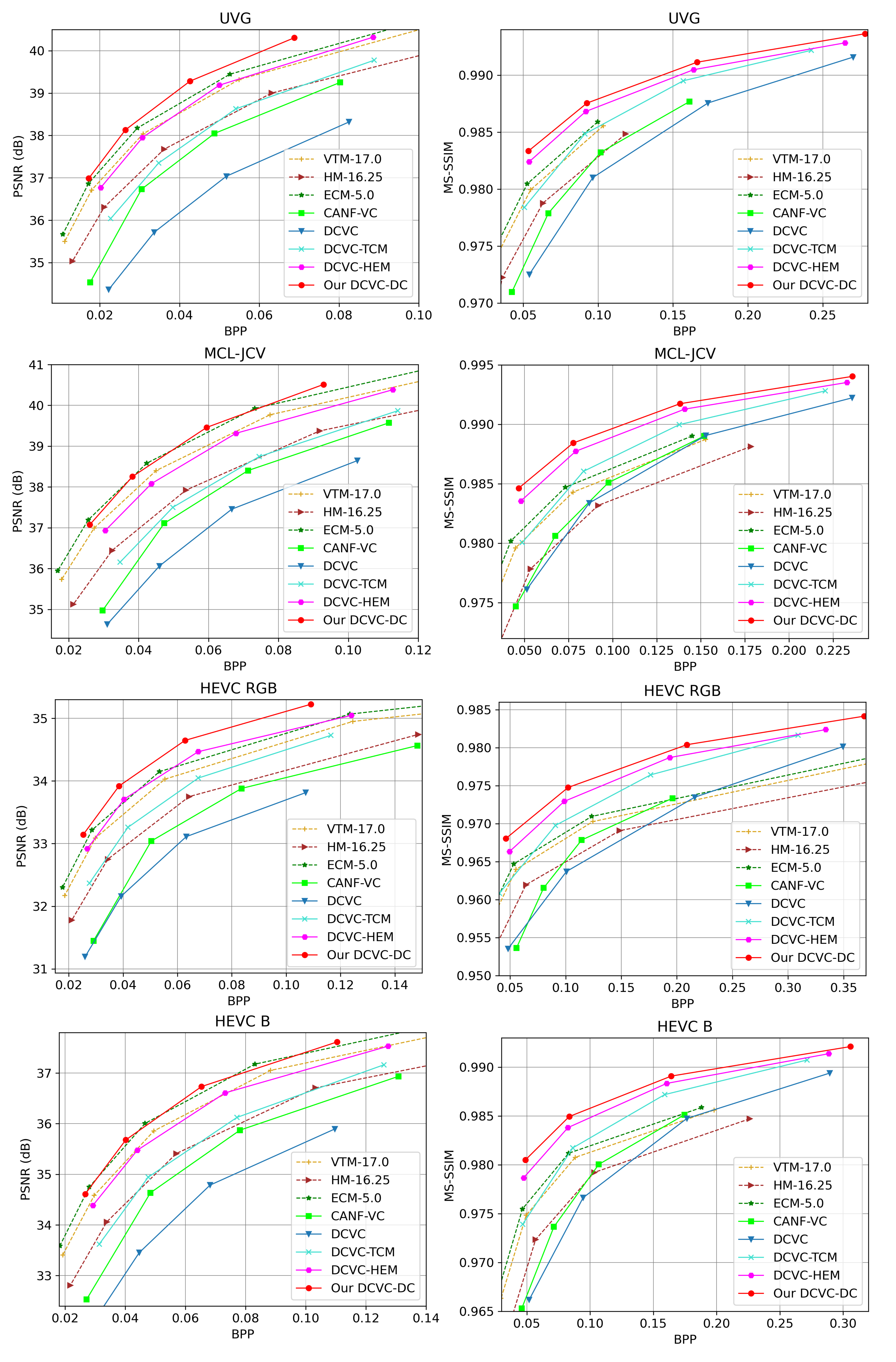}
	\end{center}
	\vspace{-0.5cm}
	\caption{RD curves of UVG, MCL-JCV, HEVC RGB and B. The comparison is in RGB colorspace with BT.709. The left column is with PSNR and right column is with MS-SSIM.	}
	\vspace{-3mm}
	\label{BT709_RGB_supp_rd_curve}
\end{figure*}

\begin{figure*}[t]
	\begin{center}
		\includegraphics[width=0.8\linewidth]{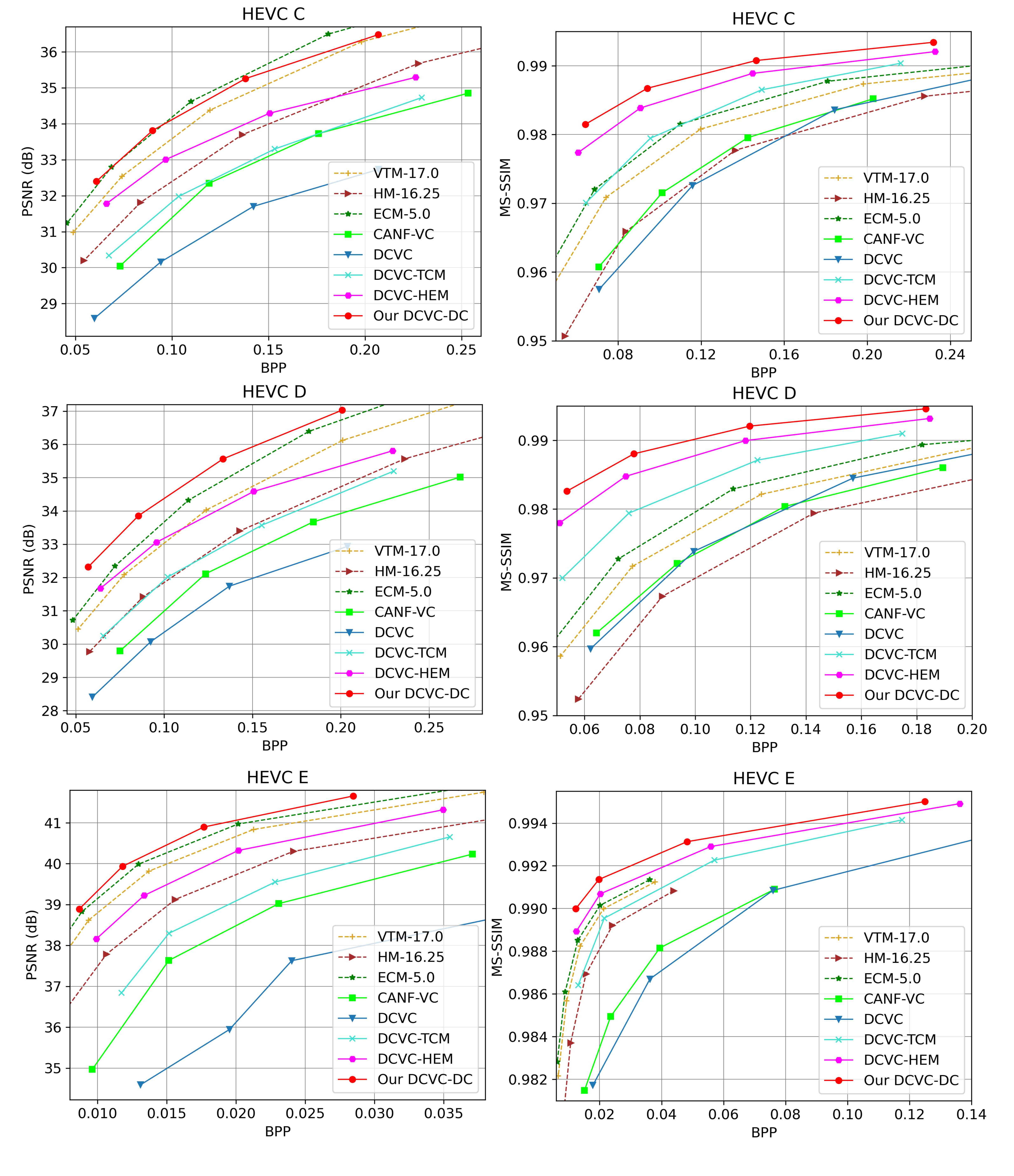}
	\end{center}
	\vspace{-0.5cm}
	\caption{RD curves of HEVC C, D, and E. The comparison is in RGB colorspace with BT.709. The left column is with PSNR and right column is with MS-SSIM.	}
	\vspace{-3mm}
	\label{BT709_RGB_supp_rd_curve2}
\end{figure*}

\begin{figure*}[t]
	\begin{center}
		\includegraphics[width=0.8\linewidth]{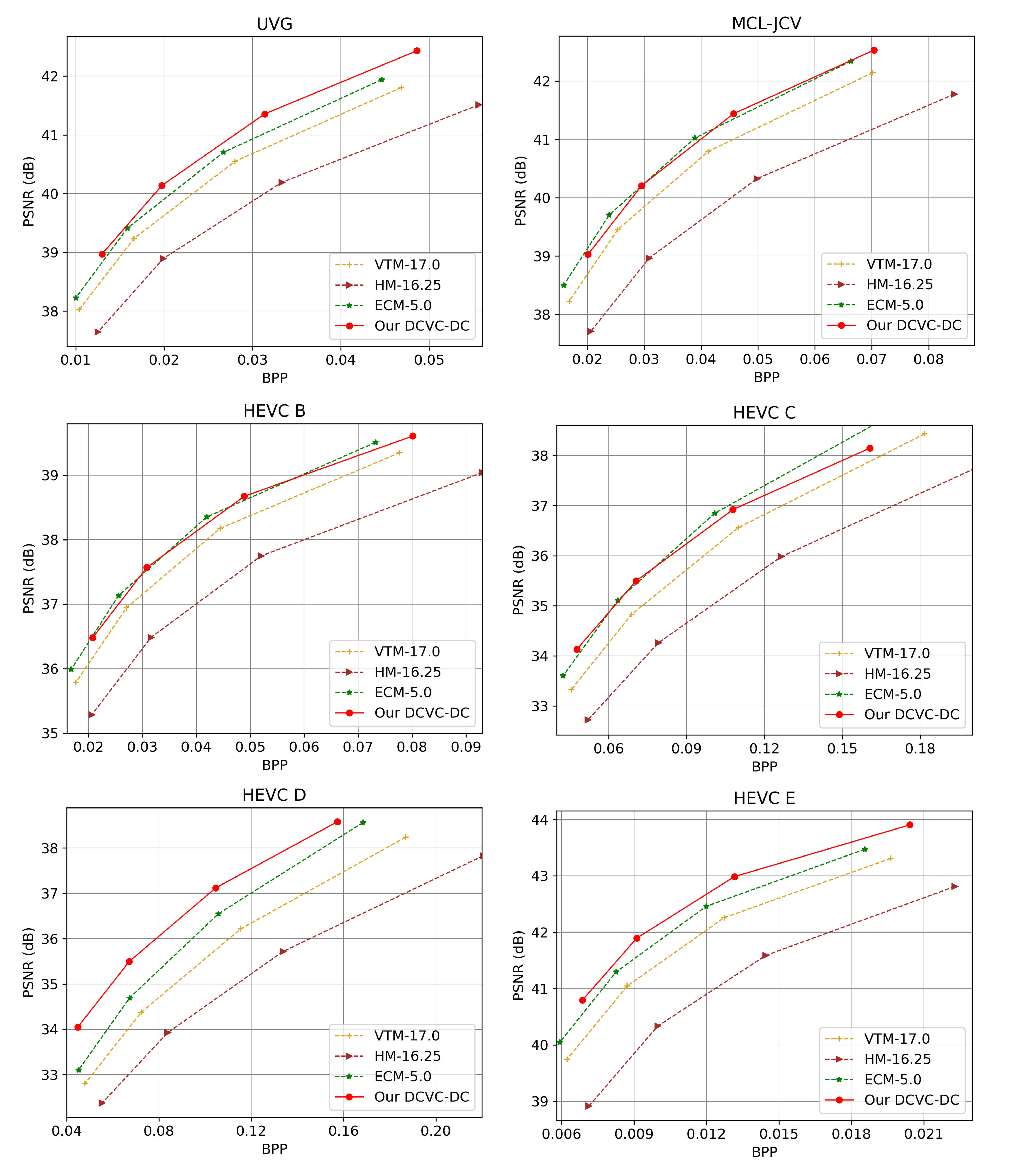}
	\end{center}
	\vspace{-0.5cm}
	\caption{RD curves of UVG, MCL-JCV, HEVC B, C, D, and E. The comparison is in YUV420. 	}
	\vspace{-3mm}
	\label{YUV420_supp_rd_curve}
\end{figure*}

In this document, we show the rate-distortion (RD) curves of all datasets, which correspond to the results in the main paper. Fig. \ref{BT709_RGB_supp_rd_curve} and \ref{BT709_RGB_supp_rd_curve2} show the RD curves for videos in RGB colorspace with  BT.709.  Fig. \ref{YUV420_supp_rd_curve} shows the RD curves for videos in YUV420 colorspace without any conversion. From these figures, we can see that our DCVC-DC can achieve SOTA compression ratio in a wide bitrate range.

\section{Visual Comparison}
\begin{figure*}[t]
	\begin{center}
		\includegraphics[width=1\linewidth]{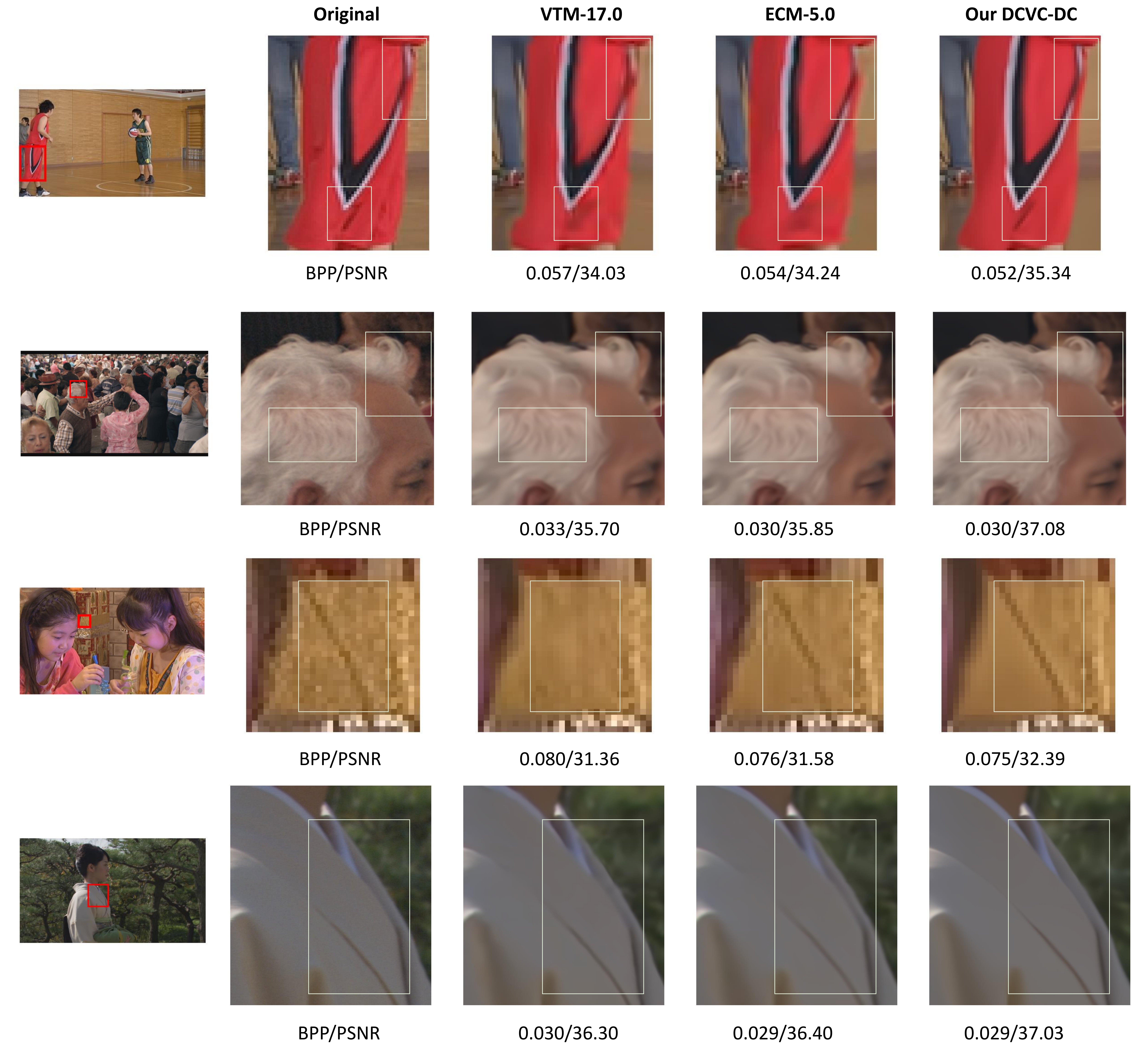}
	\end{center}
	\vspace{-0.5cm}
	\caption{Visual comparison.	}
	\vspace{-3mm}
	\label{supp_visual_example}
\end{figure*}

Here we also provide some visual comparisons to demonstrate the advantage of our  codec. Fig. \ref{supp_visual_example} shows four examples. From these examples, we can see that our DCVC-DC can reconstruct clearer textures without increasing the bitrate cost, when compared with VTM-17.0 and ECM-5.0. In addition, it is also noted that, despite we learn the hierarchical quality pattern, there is no visual flicker in the decoded video. As shown in the PSNR curve in the main paper, we can see that our DCVC-DC actually has smaller PSNR variance than  VTM-17.0. The standard community has verified the hierarchical quality pattern can improve the compression ratio with negligible visual degradation.

\end{appendices}	
	
\end{document}